\RequirePackage{fix-cm}
\documentclass[twocolumn,epjc3]{svjour3}          % twocolumn
\smartqed  % flush right qed marks, e.g. at end of proof
\usepackage{graphicx}
\usepackage{lineno}
\usepackage{multirow}
\usepackage{xcolor}
\usepackage{amssymb}
\usepackage{url}
\usepackage{textgreek}
\usepackage[normalem]{ulem}

\usepackage{amsmath}
\newcommand{\LAr}{\mbox{LAr}}

\newcommand{\ReD}{\mbox{ReD}}

\newcommand{\TPC}{\mbox{TPC}}

\newcommand{\Am}{$^{241}$Am}
\newcommand{\Kr}{$^{83m}$Kr}
\newcommand{\Qy}{$Q_{y}$}
\newcommand{\edrift}{\mathcal{E}_d}
\newcommand{\eex}{\mathcal{E}_{ex}}
\newcommand{\eel}{\mathcal{E}_{el}}

\newcommand{\Wphmax}{$W_{ph}$(max)}
\journalname{Eur. Phys. J. C}
\begin{document}

\title{Performance of the \ReD\ \TPC, a novel double-phase \LAr\ detector with Silicon Photomultiplier Readout}


%\author{none}
\author{P.~Agnes\thanksref{addr1}
  \and
 S.~Albergo\thanksref{addrunict,addrinfnct}
  \and
 I.~Albuquerque\thanksref{addusp}
  \and
 M.~Arba\thanksref{addca}
  \and
 M.~Ave\thanksref{addusp}
  \and
 A.~Boiano\thanksref{addinfnna}
 \and
 W.M.~Bonivento\thanksref{addca}
 \and
 B.~Bottino\thanksref{addinfnge,addpu}
 \and
 S.~Bussino\thanksref{adduniroma3,addinfnroma3}
\and
 M.~Cadeddu\thanksref{addca}
 \and
 A.~Caminata\thanksref{addinfnge}
 \and
 N.~Canci\thanksref{addlngs}
 \and
 G.~Cappello\thanksref{addrunict,addrinfnct}
 \and
 M.~Caravati\thanksref{addca,addunica}
 \and
 M.~Cariello\thanksref{addinfnge}
 \and
 S.~Castellano\thanksref{addinfnpi}
 \and
 S.~Catalanotti\thanksref{addinfnna,addunina}
 \and
 V.~Cataudella\thanksref{addinfnna,addunina}
 \and
 R.~Cereseto\thanksref{addinfnge}
 \and
 R.~Cesarano\thanksref{addunina}
 \and
 C.~Cical\`o\thanksref{addca}
 \and
 G.~Covone\thanksref{addunina,addinfnna}
 \and
 A.~de~Candia\thanksref{addunina,addinfnna}
 \and
 G.~De~Filippis\thanksref{addunina,addinfnna}
 \and
 G.~De~Rosa\thanksref{addunina,addinfnna}
 \and
 S.~Davini\thanksref{addinfnge}
 \and
 C.~Dionisi\thanksref{adduniroma,addinfnroma}
 \and
 G.~Dolganov\thanksref{addgrigory}
 \and
 G.~Fiorillo\thanksref{addunina,addinfnna}
 \and
 D.~Franco\thanksref{addapc}
 \and
 G.K.~Giovanetti\thanksref{addwill}
 \and
% C.~Giganti\thanksref{addlpnhe}
% \and
 C.~Galbiati\thanksref{addpu,addgssi}
 \and
 M.~Gulino\thanksref{addkore,addlns}
 \and
 V.~Ippolito\thanksref{addinfnroma}
 \and
 N.~Kemmerich\thanksref{addusp}
 \and
 I.~Kochanek\thanksref{addlngs}
 \and
 G.~Korga\thanksref{adduk}
 \and
 M.~Kuss\thanksref{addinfnpi}
 \and
 M.~La~Commara\thanksref{adduninafarmacia,addinfnna}
 \and
 L.~La~Delfa\thanksref{addca}
 \and
 M.~Leyton\thanksref{addinfnna}
 \and
 X.~Li\thanksref{addpu}
 \and
 M.~Lissia\thanksref{addca}
 \and
 S.M.~Mari\thanksref{adduniroma3,addinfnroma3}
 \and
 C.J.~Martoff\thanksref{addtemple}
 \and
 V.~Masone\thanksref{addinfnna}
\and
 G.~Matteucci\thanksref{addunina} %,addinfnna}
\and
 P.~Musico\thanksref{addinfnge}
 \and
 V.~Oleynikov\thanksref{addbudker,addnovo,addinfnna}
 \and
 M.~Pallavicini\thanksref{addunige,addinfnge}
 \and
 L.~Pandola\thanksref{addlns}
 \and
 A.~Razeto\thanksref{addlngs}
 \and
 M.~Rescigno\thanksref{addinfnroma}
% \and
% F.~Retiere\thanksref{addtriumf}
 \and
 J.~Rode\thanksref{addapc,addlpnhe}
 \and
 N.~Rossi\thanksref{addlngs}
\and
 D.~Sablone\thanksref{addlngs}
 \and
 S.~Sanfilippo\thanksref{adduniroma3,addinfnroma3}
 \and
% R.~Santorelli\thanksref{addciemat}
% \and
 E.~Scapparone\thanksref{addinfnbo}
 \and
 A.~Sosa\thanksref{addusp}
 \and
 Y.~Suvorov\thanksref{addunina,addinfnna}
 \and
 G.~Testera\thanksref{addinfnge}
 \and
 A.~Tricomi\thanksref{addrunict,addrinfnct}
 \and
 M.~Tuveri\thanksref{addca}
 \and
 M.~Wada\thanksref{addastrocent}
 \and
 H.~Wang\thanksref{adducla}
 \and
 Y.~Wang\thanksref{addihep,addihep2}
 \and
 S.~Westerdale\thanksref{addca}
}

%\authorrunning{DarkSide Collaboration} % if too long for running head
\authorrunning{\ReD\ Working Group}
%\thankstext{efai}{Funded by FAI in Naples}
\institute{Department of Physics, University of Houston, Houston, TX 7704, USA\label{addr1}
           \and
           Physics and Astronomy Department, Universit\`a degli Studi di Catania and INFN, Catania 90123, Italy\label{addrunict}
           \and
           Istituto Nazionale di Fisica Nucleare, Sezione di Catania, Catania 90123, Italy\label{addrinfnct}
           \and
           Instituto de F\'{\i}sica, Universidade de S\~{a}o Paulo, S\~{a}o Paulo 05508-090, Brasil\label{addusp}
           \and
           Istituto Nazionale di Fisica Nucleare, Sezione di Cagliari, Cagliari 09042, Italy\label{addca}
           \and
           Istituto Nazionale di Fisica Nucleare, Sezione di Napoli, Napoli 80126, Italy\label{addinfnna}
           \and          
           Istituto Nazionale di Fisica Nucleare, Sezione di Genova, Genova 16146, Italy\label{addinfnge}
           \and
           Physics Department, Princeton University, Princeton, NJ 08544, USA\label{addpu}
           \and
           Physics Department, Universit\`a di Roma3, Roma 00146, Italy\label{adduniroma3}
           \and
           Istituto Nazionale di Fisica Nucleare, Sezione di Roma3, Roma 00146, Italy\label{addinfnroma3}
           \and
           INFN Laboratori Nazionali del Gran Sasso, Assergi (AQ) 67010, Italy\label{addlngs}          
           \and
           Physics Department, Universit\`a degli Studi, Cagliari 09042, Italy\label{addunica}
           \and
           Istituto Nazionale Fisica Nucleare, Sezione di Pisa, Pisa 56127, Italy\label{addinfnpi}
           \and
           Physics Department, Universit\`a degli Studi Federico II, Napoli 80126, Italy\label{addunina}
           \and
           Physics Department, Sapienza Universit\`a di Roma, Roma 00185, Italy\label{adduniroma}
           \and
           Istituto Nazionale di Fisica Nucleare, Sezione di Roma, Roma 00185, Italy\label{addinfnroma}
           \and
           National Research Centre Kurchatov Institute, Moscow 123182, Russia\label{addgrigory}
           \and
           APC, Universit\'e Paris Diderot, CNRS/IN2P3, CEA/Irfu, Obs. de Paris, Sorbonne Paris Cit\'e , Paris 75205, France\label{addapc}
           \and
           Williams College, Physics Department, Williamstown, MA 01267 USA\label{addwill}          
           \and
           LPNHE Paris, Universit\'e Pierre et Marie Curie, Universit\'e Paris Diderot, CNRS/IN2P3, Paris 75252, France\label{addlpnhe}
           \and
           Gran Sasso Science Institute, L'Aquila AQ 67100, Italy\label{addgssi}
           \and
           Universit\`a di Enna KORE, Enna 94100, Italy\label{addkore}
           \and
           Istituto Nazionale Fisica Nucleare, Laboratori Nazionali del Sud, 95123 Catania, Italy\label{addlns}          
            \and
           Department of Physics, Royal Holloway University of London, Egham TW20 0EX, UK\label{adduk}
           \and
           Department of Pharmacy, Universit\`a degli Studi Federico II, Napoli 80131, Italy \label{adduninafarmacia}          
           \and
           Physics Department, Temple University, Philadelphia, PA 19122, USA\label{addtemple}
           \and
           Budker Institute of Nuclear Physics, Novosibirsk 630090, Russia\label{addbudker}
           \and
           Novosibirsk State University, Novosibirsk 630090, Russia\label{addnovo}
           \and
            Physics Department, Universit\`a degli Studi di Genova, Genova 16146, Italy\label{addunige}
            \and          
%           TRIUMF, 4004 Wesbrook Mall, Vancouver, BC V6T 2A3, Canada\label{addtriumf}
%           CIEMAT, Centro de Investigaciones Energ\'eticas, Medioambientales y Tecnol\'ogicas, Madrid 28040, Spain\label{addciemat}
           Istituto Nazionale di Fisica Nucleare, Sezione di Bologna, Bologna 40126, Italy\label{addinfnbo}
           \and
           AstroCeNT, Nicolaus Copernicus Astronomical Center of the Polish Academy of Sciences, 00-614 Warsaw, Poland\label{addastrocent}
           \and
           Physics and Astronomy Department, University of California, Los Angeles, CA 90095, USA\label{adducla}
           \and
           Insititute of High Energy Physics, Beijing 100049, China \label{addihep}
           \and
           University of Chinese Academy of Sciences, Beijing 100049, China \label{addihep2}            
}

\date{Received: date / Accepted: date}

\maketitle

\begin{abstract}
A double-phase argon Time Projection Chamber (TPC), with an active mass of 185\,g, has been designed and constructed %in the framework of
for the Recoil Directionality (ReD) experiment. %, a part of the DarkSide Collaboration programme.
The aim of the \ReD\ project is to investigate the directional sensitivity of argon-based \TPC s via columnar recombination to nuclear recoils in the energy range of interest (20--200\,keV$_{nr}$) for direct dark matter searches. The key novel feature of the \ReD\ \TPC\ is a readout system based on cryogenic Silicon Photomultipliers (SiPMs), which are employed and operated continuously for the first time in an argon \TPC. Over the course of six months, the \ReD\ \TPC\ was commissioned and characterised under various operating conditions using $\gamma$-ray and neutron sources, demonstrating remarkable stability of the optical sensors and reproducibility of the results. 
The scintillation gain and ionisation amplification of the \TPC\ were measured to be $g_1 = (0.194 \pm 0.013)$~photoelectrons/photon and $g_2 = (20.0 \pm 0.9)$~photoelectrons/electron, respectively.  % from fit
%$g_1 = (0.195 \pm 0.018)$~photoelectrons/photon and $g_2 = (20.7 \pm 1.6)$~photoelectrons/electron, respectively.  % from doke
%\textcolor{red}{\sout{The scintillation and ionisation response of the \TPC\ to several radioactive sources is reported and a parameterised model describing their correlation, driven by electron-ion recombination in argon, is provided for drift fields up to 1000\,V/cm.}}
The ratio of the ionisation to scintillation signals (S2/S1), instrumental for the positive identification of a candidate directional signal induced by WIMPs, has been investigated for both nuclear and electron recoils. At a drift field of 183\,V/cm, an S2/S1 dispersion of 12\% was measured for nuclear recoils of approximately 60-90\,keV$_{nr}$, as compared to 18\% for electron recoils depositing 60\,keV of energy. The detector performance reported here meets the requirements needed to achieve the principal scientific goals of the \ReD\ experiment in the search for a directional effect due to columnar recombination. A phenomenological parameterisation of the recombination probability in \LAr\ is presented and employed for modeling the dependence of scintillation quenching and charge yield on the drift field for electron recoils between 50--500\,keV and fields up to 1000\,V/cm.
  %It was found that the empirical parametrization of the recombination in Ar provided by ARIS can describe the scintillation data from ReD  up to 1000~V/cm and can be used to predict the \ReD\ ionization data for electron recoils. The parameters of the empirical function are improved by means of a global fit which consistently includes ARIS and ReD data.
    
  %\textcolor{red}{Sentences which will likely be removed}

\keywords{Time Projection Chamber \and Silicon Photomultipliers \and Noble liquid detectors}
\end{abstract}

%\linenumbers
\section{Introduction} \label{sec:intro}
Experiments searching for weakly interacting massive particles (WIMPs) play a central role in the
multifaceted effort aiming to shed light on the nature and properties of dark matter in the Universe. Of the various technologies and target materials currently used to directly detect WIMPs, liquid argon (LAr) is particularly well suited since it permits powerful background rejection via pulse shape discrimination~\cite{Amaudruz:2016dq} and additional background reduction via the use of low-radioactivity argon from underground sources~\cite{Galbiati:2007xz,Aalseth:2020nwt}. To exploit these advantages to their maximum potential, the Global Argon Dark Matter Collaboration (GADMC) is pursuing a multi-staged program to deploy a sequence of argon-based detectors that will progressively improve sensitivity to WIMPs by several orders of magnitude and ultimately reach the ``neutrino floor'', where coherent elastic neutrino interactions
%\footnote{Since these neutrino interactions are virtually indistinguishable from those induced by WIMPs, the neutrino floor represents the ultimate sensitivity at which background subtraction becomes ineffective.} 
become a significant source of background to WIMP searches. The near-term objective of the GADMC is the DarkSide-20k experiment~\cite{Aalseth:2017fik}, a double-phase argon Time Projection Chamber (TPC) currently under construction at the INFN Gran Sasso National Laboratory (LNGS). DarkSide-20k will be the first large-scale experiment to employ 1) argon extracted from underground reservoirs and 2) a readout system based on Silicon Photomultipliers (SiPMs), fulfilling two key ingredients of the GADMC program. An additional asset to the GADMC program would be to demonstrate the directional sensitivity of argon-based \TPC\ technology, since directional information provides a unique handle for discriminating against otherwise-irreducible backgrounds and is an essential requisite for correlating a candidate signal with an astrophysical phenomenon in the celestial sky~\cite{Cadeddu:2017ebu}. Hints of such directional sensitivity have been observed by the SCENE experiment~\cite{Cao:2014gns}.

Here we report on the operation and characterisation of a double-phase argon \TPC\ developed for the Recoil Directionality (\ReD) experiment, a part of the programme pursued by the DarkSide Collaboration. The main scientific goal of the \ReD\ project is to investigate the directional sensitivity and performance of \LAr\ \TPC\ technology in the energy range of interest for WIMP-induced nuclear (Ar) recoils (20--200\,keV$_{nr}$). 
\begin{figure}[tbp]
	\centering 
	\includegraphics[width=.85\columnwidth]{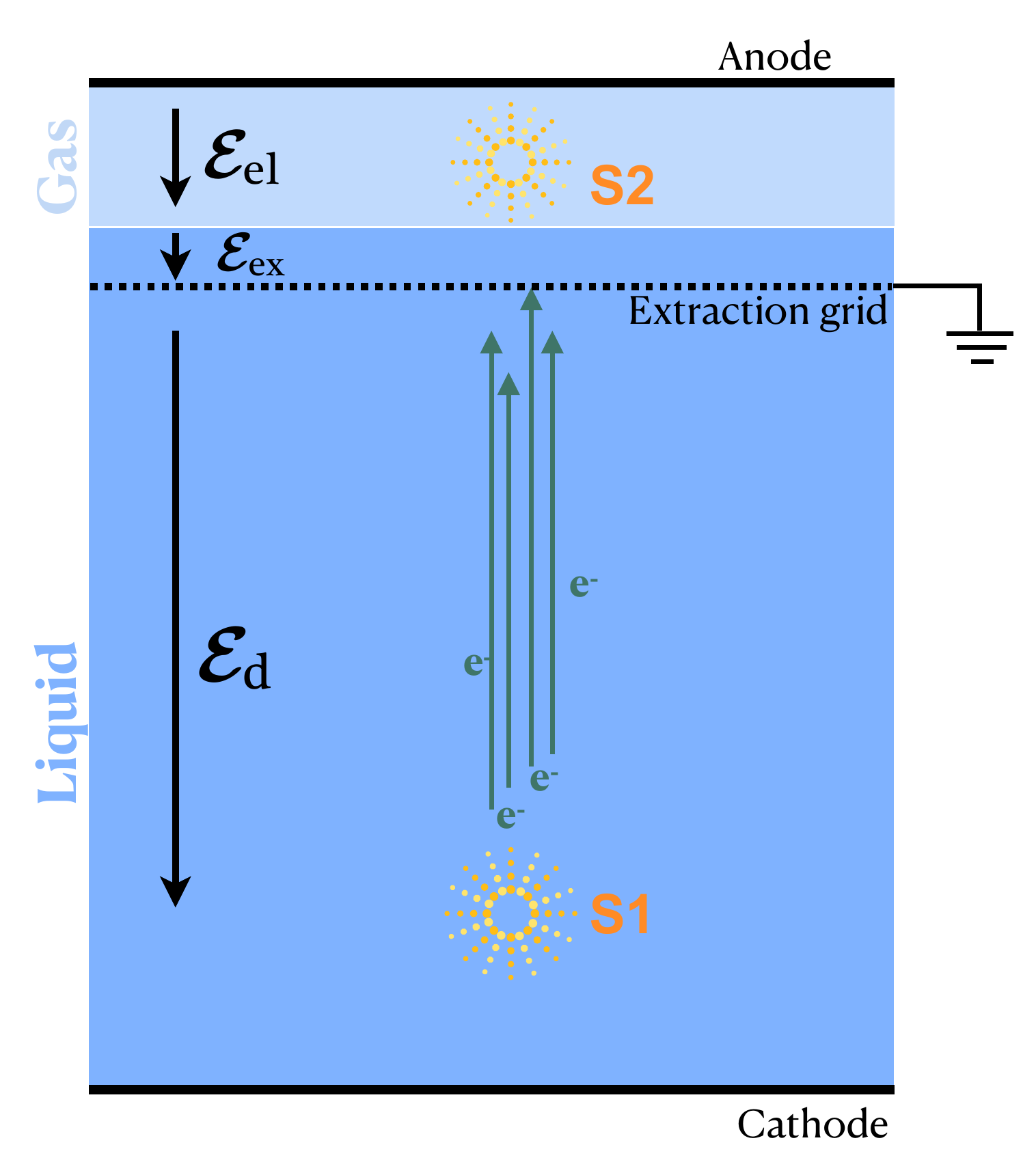}
	\caption{\label{fig:generalTPC} Conceptual sketch of the working principle of a double-phase argon \TPC, illustrating a typical event with both primary scintillation (S1) and secondary ionisation (S2) signals. The electric field in three different regions of the \TPC, indicated here as the drift field ($\edrift$), extraction field ($\eex$) and electroluminescence field ($\eel$), is responsible for drifting free ionisation electrons towards the extraction grid, extracting them into the gas phase, and producing the electroluminescence signal in the gas, respectively. }
\end{figure}
The \ReD\ \TPC, like any generic double-phase noble liquid \TPC, consists of a volume filled with a liquid target, above which lies a thin layer of the same element in the gaseous phase (the ``gas pocket'') in equilibrium with the layer below, as illustrated in Fig.\,\ref{fig:generalTPC}. When an ionising particle deposits energy in the liquid volume, target atoms are excited and ionised, creating excitons and electron-ion pairs. The excitons emit scintillation light, as do a fraction of the electron-ion pairs that recombine; this signal is referred to as S1. The electrons surviving recombination are then drifted towards the liquid-gas boundary, extracted into the gas phase, and accelerated in the gas pocket by a suitable electric field with magnitudes $\edrift$ (drift field), $\eex$ (extraction field), and $\eel$ (electroluminescence field), respectively. The acceleration of electrons by the electric field in the gas pocket produces a secondary signal composed of electroluminescent light~\cite{Buzulutskov:2020xhd} and referred to as S2, whose amplitude is proportional to the number of electrons surviving recombination and whose delay with respect to S1 is equal to the time needed for the electrons to drift through the liquid. Both S1 and S2 signals are detected by photosensors externally viewing the active volume. 
%The electrons surviving recombination are drifted up to the liquid-gas boundary by a suitable electric field (drift field, $\edrift$), extracted into the gas phase with a stronger electric field (extraction field, $\eex$), then accelerated in the gas pocket by the electroluminescence field ($\eel$) to produce a secondary light signal, called S2. 
%The amplitude of S2 is proportional to the number of electrons surviving recombination and delayed by the time needed for the electrons to drift through the liquid. 

%Two key features of double-phase TPCs make them particularly appealing for WIMP dark matter searches: 1) their 3D sensitivity to the interaction point (the $x$-$y$ coordinates are inferred by the reconstructed position of the S2 interaction, while the $z$ coordinate is obtained from the delay time between S1 and S2 signals); and 2) their capability to discriminate nuclear recoils (NRs) from electron recoils (ERs) via a combined measurement of S1 and S2. In the case of argon TPCs, the time structure of the S1 signal provides an additional, powerful way to discriminate between ERs and NRs, a technique known as pulse shape discrimination~\cite{Amaudruz:2016dq}.

%in YB etc. would do (Astroparticle Physics 85 (2016) 1–23).
%
%The latest TDR cite instead DEAL3600 for PSD:
%
%P. A. Amaudruz et al. (The DEAP-3600 Collaboration), Phys. Rev. Lett. 121, 071801 (2018).
%[19] R. Ajaj et al. (The DEAP Collaboration), Phys. Rev. D 100, 022004 (2019).
%
%But with SiPM....
%

The potential directional sensitivity of a double-phase \TPC\ stems from the dependence of columnar recombination on the alignment of the recoil momentum with respect to the drift field~\cite{Cataudella:2017kcf}. To further investigate this process, the \ReD\ detector was irradiated with neutrons of known energy and direction, produced via p($^7$Li,$^7$Be)n by the TANDEM accelerator at the INFN LNS laboratory in Catania. Monte Carlo (MC) simulations were used to define and benchmark detector performance requirements needed to conclusively identify a directional effect with a size as reported by the SCENE collaboration.
They include: scintillation gain ($g_{1}$) greater than approximately 0.2 photoelectrons/photon; ionisation amplification ($g_{2}$) greater than 15 photoelectrons/electron; and S2/S1 dispersion better than approximately 10-15\% for nuclear recoils with 70\,keV$_{nr}$. These performance specifications are measured and evaluated with the \ReD\ detector discussed here.

The \ReD\ \TPC\ shares several key characteristics of the future DarkSide-20k experiment, including some elements of the mechanical design, but on a smaller scale. The main technological advance is a readout system based entirely on SiPMs, which offer the possibility of a higher photon detection efficiency relative to typical cryogenic photomultipliers~\cite{Aalseth:2017fik,Rossi_2016,Aalseth:2020zdm}.
%a significantly higher photon detection efficiency than the cryogenic photomultipliers used in the predecessor DarkSide-50 experiment \textbf{[ref]}. 
The \ReD\ detector discussed here is the first double-phase argon \TPC\ read out by SiPMs to be stably operated on the time scale of months and fully characterised with neutron and $\gamma$ sources.
%This work thus represents an important benchmark test for DarkSide-20k.
In the future, the \ReD\ \TPC\ could also be used to characterise the response of the detector to very low-energy nuclear recoils (below a few keV), a potential signature of light WIMPs with masses of a few GeV/c$^2$, as studied by the predecessor DarkSide-50 experiment~\cite{Agnes:2015gu,Agnes:2015ftt,Agnes:2018ves}. Measurements by \ReD\ could be used to improve the understanding of ionisation models for future searches in such uncharted regimes.

The \ReD\ \TPC\ data reported here were taken 
%This paper reports on data taken with the \ReD\ detector 
at the INFN ``CryoLab'' at the University of Naples Federico II, while operating continuously between 7 June 2019 and 18 November 2019 (165 days). This paper is organised as follows: the \ReD\ \TPC\ and the experimental setup are described in Sec.\,\ref{sec:detector}. Sec.\,\ref{sec:ser} reports on the response of the SiPMs to single photons and the corresponding calibration procedure. The performance and response of the \TPC\ to scintillation light is described in Sec.\,\ref{sec:s1}, while Sec.\,\ref{sec:s2} reports on the characterisation of the combined scintillation-ionisation response. The dependence of S1 and S2 on the drift field, for $\edrift$ up to 1000\,V/cm, is discussed in Sec.\,\ref{sec:s1s2}. Conclusions and a summary are presented in Sec.\,\ref{sec:conclusions}.

\section{Experimental setup} \label{sec:detector}
The experimental setup includes the \TPC\ and also the readout, data acquisition, cryogenic and control systems needed to operate the complete system. Here we give a detailed description of each component.
\subsection{TPC}
\label{sec:tpc}
\begin{figure*}[htbp]
	\centering % \begin{center}/\end{center} takes some additional vertical space
	\includegraphics[height=.42\textwidth]{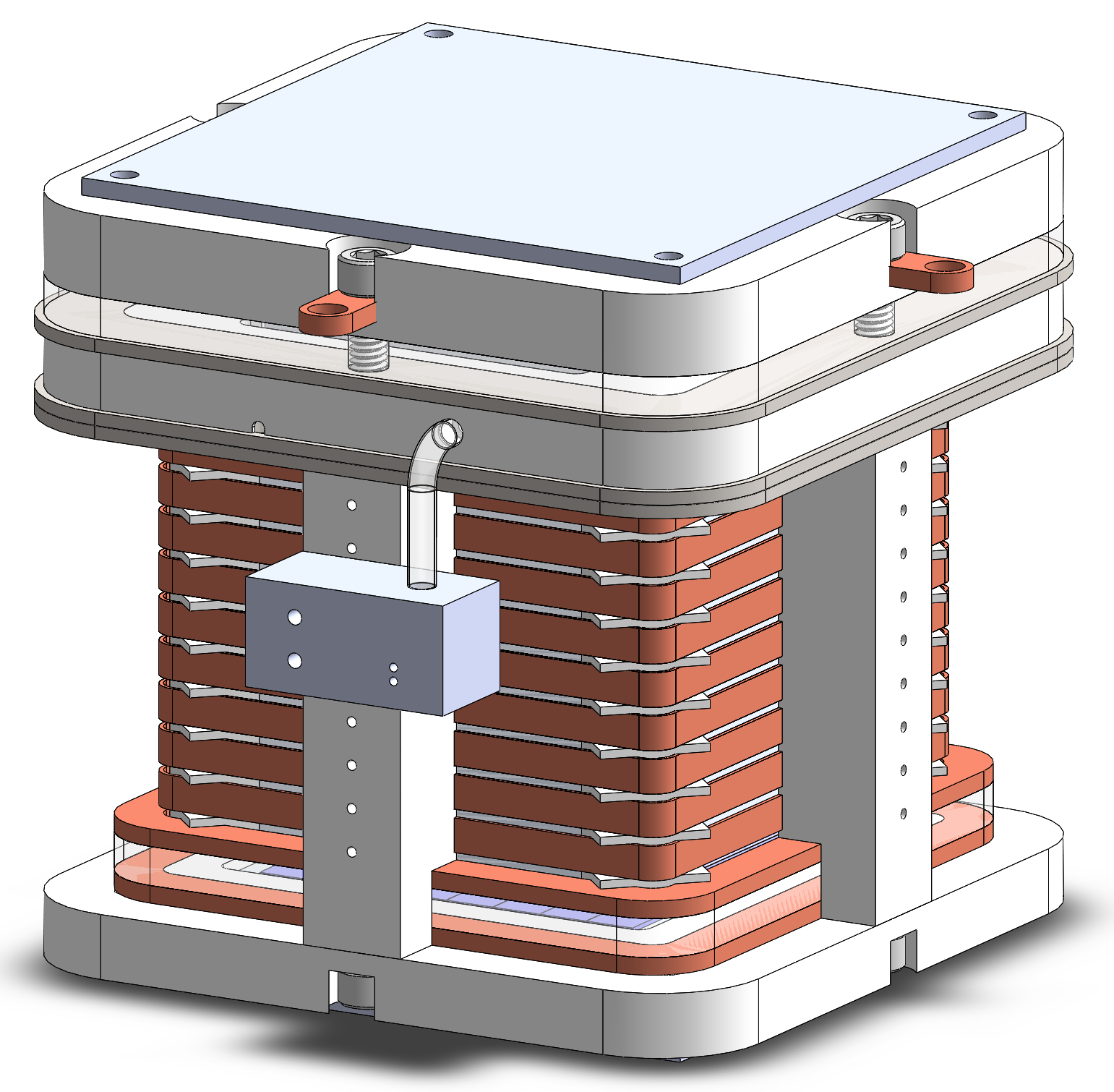}
	\qquad
	\hspace{0.2cm}
	\includegraphics[height=.42\textwidth,origin=c]{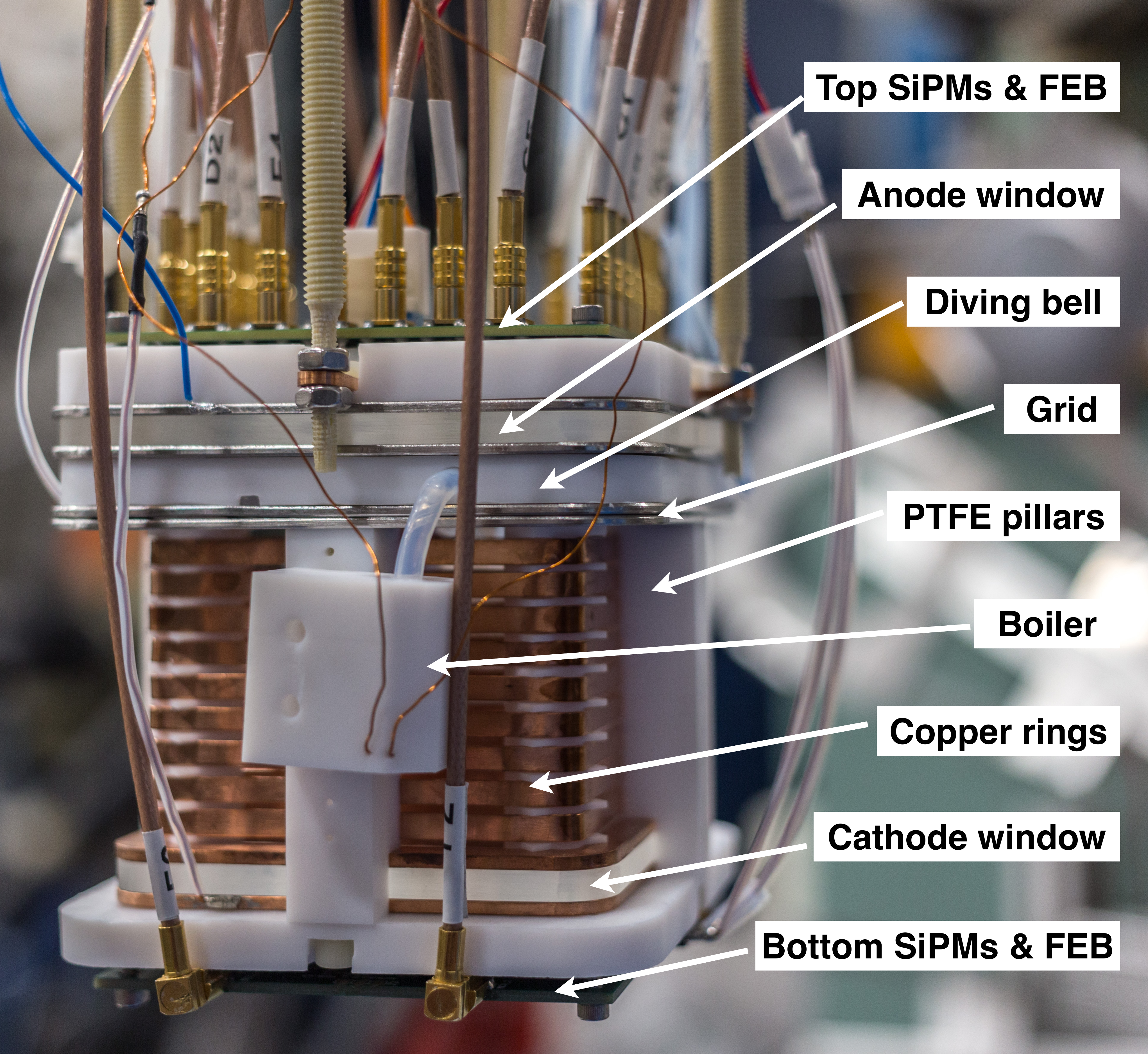}
	% "\includegraphics" from the "graphicx" permits to crop (trim+clip)
	% and rotate (angle) and image (and much more)
	\caption{\label{fig:tpc} A drawing (\textit{left}) and picture (\textit{right}) of the \ReD\ \TPC.}
\end{figure*}

The heart of the \ReD\ experiment is the \TPC, illustrated in Fig.\,\ref{fig:tpc}.
%''Physical description''
The active volume is a cuboid of 5\,cm\,$\times$ 5\,cm\,$\times$\,6\,cm (\textit{l}\,$\times$\,\textit{w}\,$\times$\,\textit{h}). It is delimited by two 4.5-mm-thick acrylic windows on top and bottom, while the side walls are composed of two 1.5-mm-thick acrylic plates, interleaved with 3M\texttrademark\ enhanced specular reflector foil. The top and bottom windows are covered on both sides by a 25-nm-thick transparent conductive layer of indium tin oxide (ITO), which allows the windows to serve as electrodes (anode and cathode) via the application of an electric potential. 
The extraction grid is made of 50-\textmu m-thick stainless steel etched mesh, with 1.95-mm-wide hexagonal cells and an optical transparency of 95\%. It is located 10\,mm below the top acrylic window.
All \TPC\ parts are locked together by eight PTFE pillars and two PTFE square frames (top and bottom) that host the photosensors.
A third PTFE frame for the gas pocket is inserted between the anode window and the grid. Since the photons emitted by argon scintillation have a wavelength of 128\,nm, outside the sensitive region of typical photosensors, they are first converted into visible light for detection. The internal surfaces of the \TPC\ are therefore fully covered with a ($180 \pm 20$)\,\textmu g/cm$^2$ layer of tetraphenyl butadiene (TPB), which acts as a wavelength shifter and re-emits photons at $\sim$420\,nm, to which the photosensors have optimal sensitivity.

%Single-vs. double phase
The \TPC\ can be operated in single-phase mode, as a scintillation-only detector with the inner volume entirely filled with $\sim$185\,g of LAr, or alternatively in double-phase mode, with the additional presence of a gas pocket. The gas pocket is created by a boiler that consists of a platinum resistance temperature sensor (Pt-1000 RTD) acting as a heater, enclosed in a PTFE box located on one of the external pillars and powered at 20\,V. The height of the gas pocket is mechanically fixed at $\sim$7 mm due to a hole located in one of the side walls above the grid. When operated in double-phase mode, the \TPC\ has a maximum drift length of 50\,mm (equal to the distance between the cathode and grid), in addition to a ($3 \pm 1$)-mm-thick LAr layer above the grid and a ($7 \mp 1$)-mm-thick gas pocket.

%Electric fields
The electric field within the \TPC\ is generated by applying voltage independently to the anode and cathode using a CAEN N1470 HV power supply module, while keeping the extraction grid at ground. To improve the uniformity of the field in the drift region, the \TPC\ walls are externally surrounded by a field cage composed of nine horizontal copper rings connected via a chain of resistors and spaced 0.5\,cm apart. Voltage is applied independently to the first ring, closest to the grid. The electric field differs in magnitude between the extraction region above the grid and the gas pocket due to the different dielectric constants of liquid and gaseous argon, taken here as $\epsilon_{l} = 1.505$ and $\epsilon_{g}=1.001$~\cite{Amey1964}, respectively.
%The electric fields required to drift the electrons in the liquid, extract them from the liquid to the gas, and produce the electroluminescence signal in the gas are created by applying independent voltages to the anode and cathode using a CAEN N1470 HV power supply module. The drift field ($\edrift$) in the liquid is established between cathode and grid, which is always kept at ground. In order to improve the uniformity of this field, the \TPC\ walls are externally surrounded by a field cage composed of nine horizontal copper rings spaced 0.5\,cm apart, the first of which (closest to the grid) can be given an independent voltage, as discussed below. The potential difference between the grid and the anode produces two electric fields, one in liquid and one in gas, of different magnitudes due to the difference in dielectric constants of liquid and gaseous argon, $\epsilon_{l} = 1.505$ and $\epsilon_{g}=1.001$~\cite{Amey1964}, respectively. The extraction field ($\eex$), present in the liquid above the grid, must be strong enough to allow the drifted electrons to pass into the gaseous phase. These electrons are then accelerated by the electroluminescence field ($\eel$) within the gas pocket in order to produce the secondary signal. 

\begin{figure}[htbp]
	\centering % \begin{center}/\end{center} takes some additional vertical space
	\includegraphics[width=.90\columnwidth]{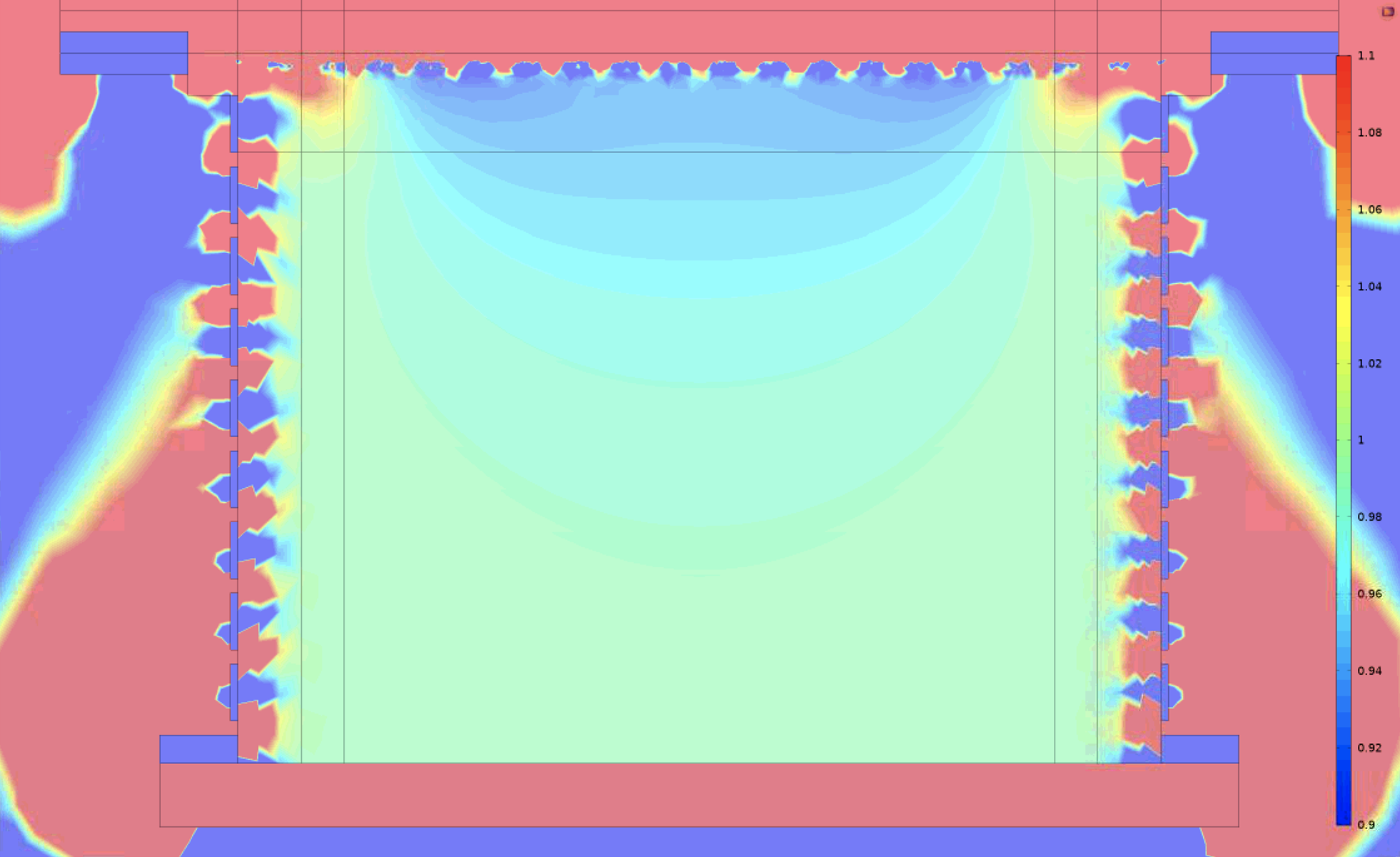}
	\caption{\label{fig:Efield_map} Relative electric field map within the \TPC\ for the reference voltage configuration: $+5211$\,V (anode), $+86$\,V (first ring) and $-744$\,V (cathode). The colour scale is set to $\pm$10\% relative to $\langle \edrift \rangle = 183$\,V/cm. Figure made with COMSOL\textregistered~\cite{COMSOL}.}
\end{figure}

% MR changed following Wlter comments
Voltage settings corresponding to $100 \leq \edrift \leq 1000$\,V/cm were originally selected by means of a simplified simulation of the \ReD\ \TPC\ in COMSOL\textregistered~\cite{COMSOL}, while actual average field values in each of the three regions were calculated \textit{a posteriori} using a fully detailed model of the detector including the three-dimensional structure of the mesh.  
%Voltage configurations corresponding to $100 \leq \edrift \leq 1000$\,V/cm were chosen by means of a simplified 2D simulation in COMSOL\textregistered~\cite{COMSOL}, while average field values in each of the three regions were calculated \textit{a posteriori} using a more detailed 3D simulation. 
The difference in the calculated drift field between the simplified and full simulations is negligible at high fields and becomes progressively larger for lower fields, reaching up to 20\% at the lowest studied value of 100\,V/cm. For the reference voltage configuration discussed in this work, $+5211$\,V (anode), $+86$\,V (first ring) and $-744$\,V (cathode), a drift field of $\sim$200\,V/cm was obtained using the simplified COMSOL simulation, while the full simulation calculated $\langle \edrift \rangle = (183 \pm 2)$\,V/cm in the inner part of the \TPC. Fig.\,\ref{fig:Efield_map} shows a map of the electric field within the \TPC\ for the reference voltage configuration, relative to the average value of 183\,V/cm in the drift region. Field inhomogeneities were found to be significant (up to 20\%) only for low-field configurations. For all double-phase data presented here, the extraction and electroluminescence fields were calculated as $\langle \eex \rangle=(3.8 \pm 0.2$)\,kV/cm and $\langle \eel \rangle =(5.7 \pm 0.2)$\,kV/cm, respectively, sufficiently strong to fully extract all electrons from the liquid to gas phase~\cite{Chepel:2012sj}. Since $\eex$ and $\eel$ are determined mostly by the anode voltage and the thickness of the gas pocket, the difference in the calculated values between the simplified and full simulations was found to be below 3\%.

\subsection{Silicon photomultipliers and readout system}

Customised NUV-HD-Cryo SiPMs~\cite{Gola:2019idb}, developed specifically for the DarkSide project by Fondazione Bruno Kessler (FBK), are used to detect the light signals in the  \ReD\ \TPC. They have a maximum photon detection efficiency at $\sim$420\,nm ($>50$\% at room temperature) and a high-density distribution of Single Photon Avalanche Diodes (SPADs)~\cite{sipmnuv:2017}. The \ReD\ SiPMs are characterised by a triple doping concentration, 25-\textmu m cell pitch and 10 M\textOmega\ quenching resistance at cryogenic temperature. Each SiPM measures 11.7\,mm\,$\times$\,7.9\,mm and is assembled onto a 5\,cm\,$\times$\,5\,cm tile with 24 devices.
%Each rectangular SiPM measures 11.7$\times$7.9 mm$^2$ and is assembled into one of two square tiles (5\,cm $\times$ 5\,cm), each containing 24 devices.
%In particular every SiPM is
%glued on an arlon substrate and connected using the wire bonding technique. 
One tile is positioned at the top of the \TPC\ and one tile at the bottom,  
behind their respective acrylic windows. Since the position of an S2 event in the gas pocket provides a reasonable estimate of the $x$-$y$ coordinate of the corresponding interaction point in the \TPC, the 24 SiPMs of the top tile are read out individually for improved spatial resolution, while those of the bottom tile are summed in groups of six.
%As the position of the  S2 signal in the gas pocket is a proxy for the $x-y$ coordinates of the events in the \TPC, the 24 SiPMs of the  top tile are read out individually for improved resolution, while those of the bottom tile are summed in groups of six. 
%For improved $x-y$ position resolution of the S2 signal, the 24 SiPMs of the top tile are read out individually, 

\begin{figure*}[htbp]
	\centering % \begin{center}/\end{center} takes some additional vertical space
	\includegraphics[height=.40\textwidth]{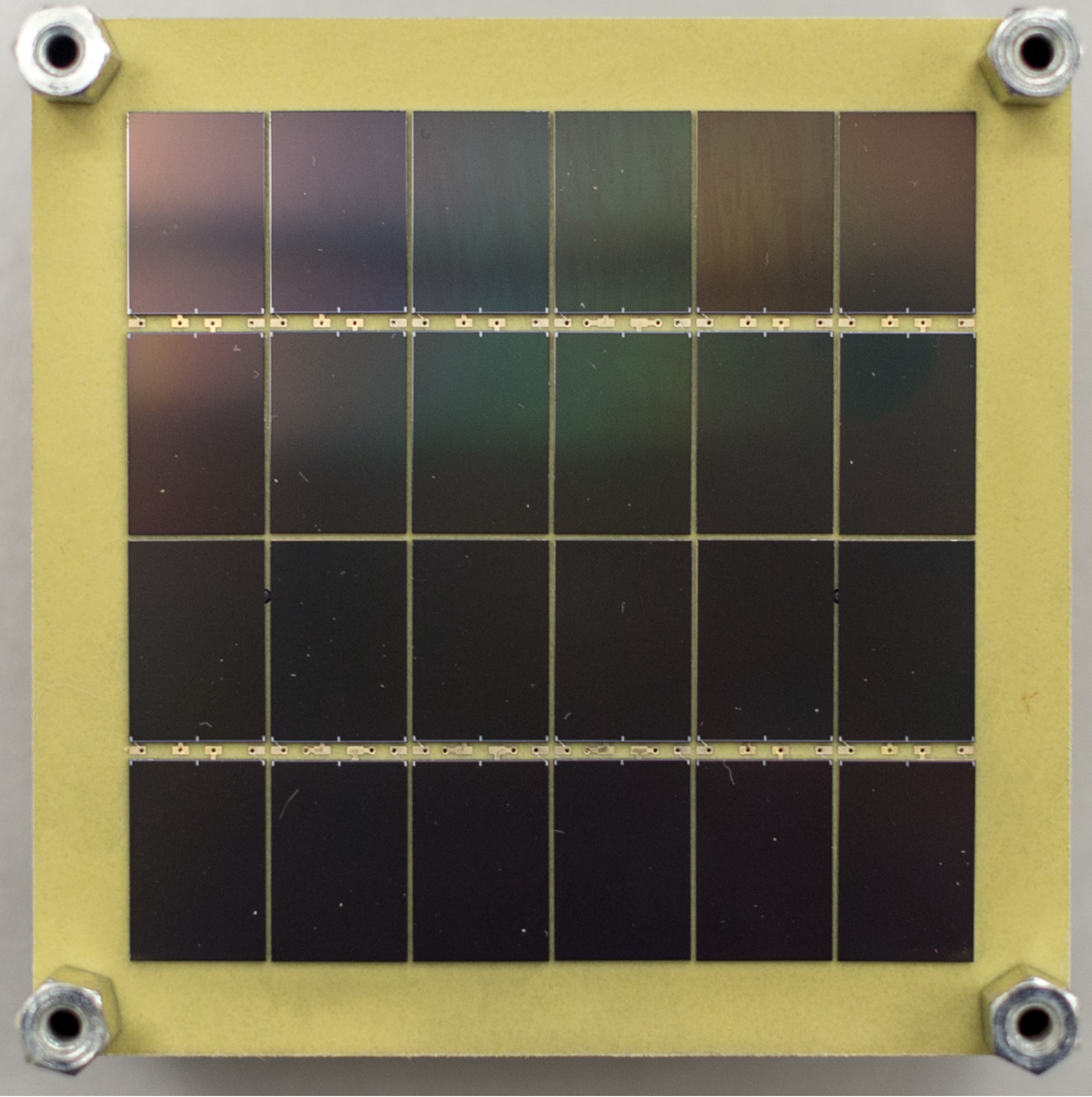}
	\hspace{0.7cm}
	\includegraphics[height=.40\textwidth]{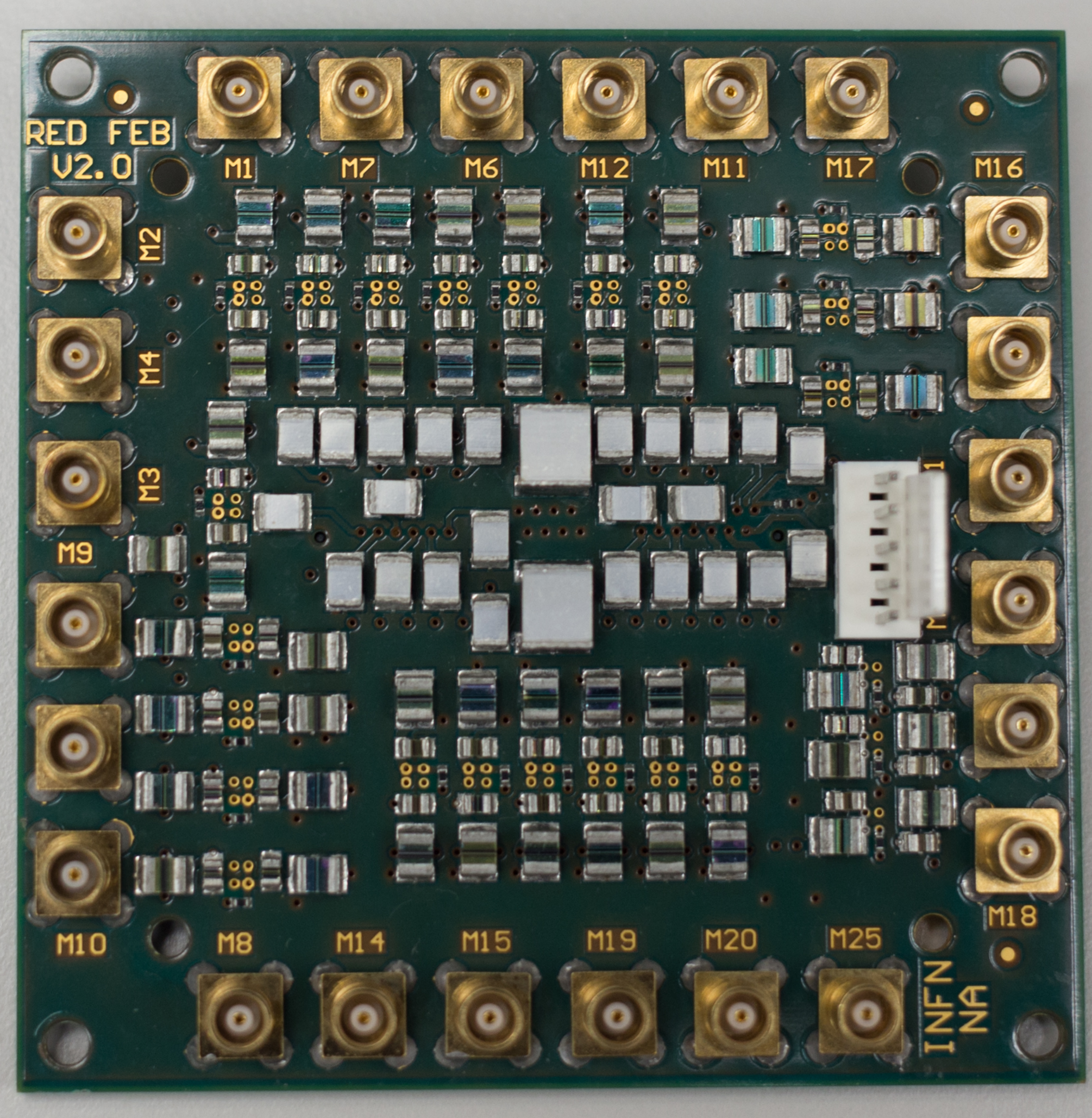}
	%\qquad
	%\includegraphics[width=.45\textwidth,origin=c]{figure/feb24front}
	% "\includegraphics" from the "graphicx" permits to crop (trim+clip)
	% and rotate (angle) and image (and much more)
	\caption{\label{fig:feb} \textit{Left:} 24 rectangular NUV-HD-Cryo SiPMs from FBK, assembled onto a 5\,cm\,$\times$\,5\,cm tile.  \textit{Right:} the 24-channel Front-End Board (FEB) used for reading out the top tile. }
\end{figure*}

The Front-End Boards (FEBs) supply power to the photosensors and amplify the output signals at cryogenic temperature. Each SiPM is operated at a fixed bias voltage of 34\,V, corresponding to 7\,V of overvoltage with respect to the breakdown voltage. The FEBs employ a low-noise amplifier~\cite{coldpreamp:2018} developed specifically to ensure optimal performance of the device at its normal working temperature of $\sim$87 K and whose design is based on a high-speed, ultra-low-noise operational amplifier (LMH6629SD) from Texas Instruments.  Due to the differing readout schemes, the FEBs for the top and bottom tiles are distinct. The top FEB handles each SiPM independently: the common bias voltage is distributed to the 24 devices and the signals are read and amplified one-by-one. The bottom FEB operates four quadrants, each made of three branches with two SiPMs in series~\cite{tile:2018}. 68\,V ($2 \times 34$\,V) are distributed to each of the three branches and the six SiPM signals are summed and amplified, giving a total of four output channels. Fig.\,\ref{fig:feb} shows the top tile with 24 SiPMs and the corresponding 24-channel FEB.
%~\footnote{The custom-made top FEB was developed and provided by the INFN Naples.}.
%The single channel reading guarantees high granularity, useful for studying the signal inside the gas
%pocket. In particular this improves the XY position and the energy reconstruction. The PCB board is realized on vetronite FR4 material and developed in four layers. Due to the low 24 amplifying channels
%input impedance, an internal ground layer has been placed; moreover, the SiPM bias supply filtering and distribution has been 
%very carefully designed and realized.\\

\subsection{Data acquisition}
%---
The data acquisition system for the \ReD\ \TPC\ is based on two CAEN V1730 Flash ADC Waveform Digitizers, for a total
of 32 acquisition channels, of which 28 are used for reading out the SiPMs. Signals are digitised with 14-bit resolution at a sampling rate of 500\,MHz. Upon receiving a trigger, data are saved in internal buffers, then asynchronously transferred to a Linux data acquisition server via an optical link connected to an A2818 CAEN PCI controller. The software was built upon a package developed for the PADME experiment~\cite{Leonardi_2017} and is based on the CAEN Digitizer Libraries. A centralised server controls several independent readout processes, one for each board.
%The user interact with the
%server via a python graphical user interface.
Event building is performed offline at the event reconstruction stage. 
% text from E.Leonardi
%The PadmeDAQ package was developed by the PADME collaboration. It is based on the CAENDigitizer libraries and implements a
%full interface to the V1742 board functionalities. All module activities can be easily controlled with a human readable
%configuration file.
The trigger is implemented via the logic of the digitizer boards and generally consists of (at least) two independent signals from the logical \texttt{OR} of the two pairs of readout channels on the bottom tile. The default trigger condition requires a coincidence within 200\,ns of these two trigger signals with an individual threshold set approximately at 2 photoelectrons (PE). This trigger is fully efficient for signals above 100 PE.

\subsection{Cryogenic system and control infrastructure}\label{sec:cryo}

The \ReD\ detector has a dedicated cryogenic system designed to liquify and continuously purify evaporated argon gas from the cryostat. The cryostat is first filled with research-grade gaseous argon (N6.0), which is then cooled by means of a Cryomech PT90 cold head. 
%The cryogenic system consists of the following key elements: the custom-made cryostat and condenser, the gas
%recirculation dry pump, the purification getter and
%the gas panel (a sub system devoted to the control and management of the argon flow).
%The entire assembly, which is shown in Fig.~\ref{fig:crio} is mounted on a movable cart integrated with a lifting mechanism allowing to variate the
%height of the cryostat with respect to the ground.
%The filling of \LAr\ is performed using research-grade gaseous argon (6.0), w
%After the pump-and-purge cycles of the initial purification phase, the argon gas, at room temperature, is sent to the hot getter for the first purification, 
%then to the condenser for liquefaction and then, in liquid form, by gravity it reaches the cryostat.
%The cooling power in the condenser is provided by a Cryomech PT90 cold head~\cite{compressor}.
%coupled with the custom-made copper flange designed with the goal of maximizing the contact surface between the cold metal and the arriving warm argon gas.
After the initial cooling of the TPC and cryostat, the LAr level is monitored by means of two Pt-1000 RTDs mounted inside the cryostat. Given the thermal load of the system, the filling procedure, starting from the aperture of the Ar gas bottle to the accumulation
of $\sim$30\,cm of LAr takes approximately 12 hours.
%\begin{figure}[htbp]
%	\centering % \begin{center}/\end{center} takes some additional vertical space
%	\includegraphics[width=.6\textwidth]{figure/ReD_System_rev}
%	% "\includegraphics" from the "graphicx" permits to crop (trim+clip)
%	% and rotate (angle) and image (and much more)
%	\caption{\label{fig:crio} Photo of the \ReD\ assembly in Naples. The TPC is deployed inside the cryostat, which is
%       filled by the condenser. The entire system is mounted on a movable cart.}
%\end{figure}
Once the cryostat is filled, the argon gas source is excluded and the system is switched to recirculation mode. The system then operates in a closed loop: the argon from the cryostat ullage is extracted by a dry pump, pushed through a SAES hot getter for purification and finally re-condensed back into the cryostat. 

All detectors and sensors described above can be operated and read out remotely by means of a slow control system that allows a user to perform basic operations via a graphical user interface (GUI), e.g.\ enabling/disabling the voltage delivered by an instrument. By connecting each individual component to a NI-PXIe-8840 controller, the slow control system continuously monitors all operating parameters and stores them in a database. The slow control software is written in LabVIEW~\cite{LabView}, with each instrument piloted by its own stand-alone application.

\section{Photosensors and single-photoelectron response} \label{sec:ser}
The Single photoElectron Response (SER) of the \ReD\ \TPC\ is studied using a 
Hamamatsu PLP-10 pulsed diode laser with a wavelength of 403\,nm, externally triggered at 100\,Hz. Pulse emissions of 50\,ps are delivered to the inner volume of the \TPC\ via optical fibres and the signal responses from each of the 28 SiPM readout channels are digitised inside an acquisition window of 20\,\textmu s. Laser calibrations were regularly performed during the 165~days of continuous operation of the system. 
%The integral of the voltage signal,
%which is referred as ``charge'' in the following, is a quantitative measure of the response
%of the SiPMs.

The charge measured by each SiPM is calculated offline by integrating the digitised waveform, following subtraction of the average baseline, over a fixed window of 4\,\textmu s starting 600\,ns before the external trigger time. Data are also processed by applying a more CPU-intensive digital filtering technique that allows the counting of single photoelectrons. The filter deconvolves the response function of the SiPM %(see App.\,\ref{sec:sipmtime})
and the filtered signal is then scanned for photoelectron peaks. The peak identification is performed by the \texttt{find\_peaks} algorithm from the \texttt{SciPy} libraries~\cite{2020SciPy-NMeth,SciPy};
the algorithm also estimates the peak ``prominence'', defined as the height of the filtered peak. The total prominence summed over all peaks is then taken as a proxy of the number of detected photoelectrons. 

\begin{figure}
\centering
\includegraphics[trim=0 0 1.4cm 0.5cm,clip,width=0.95\columnwidth]{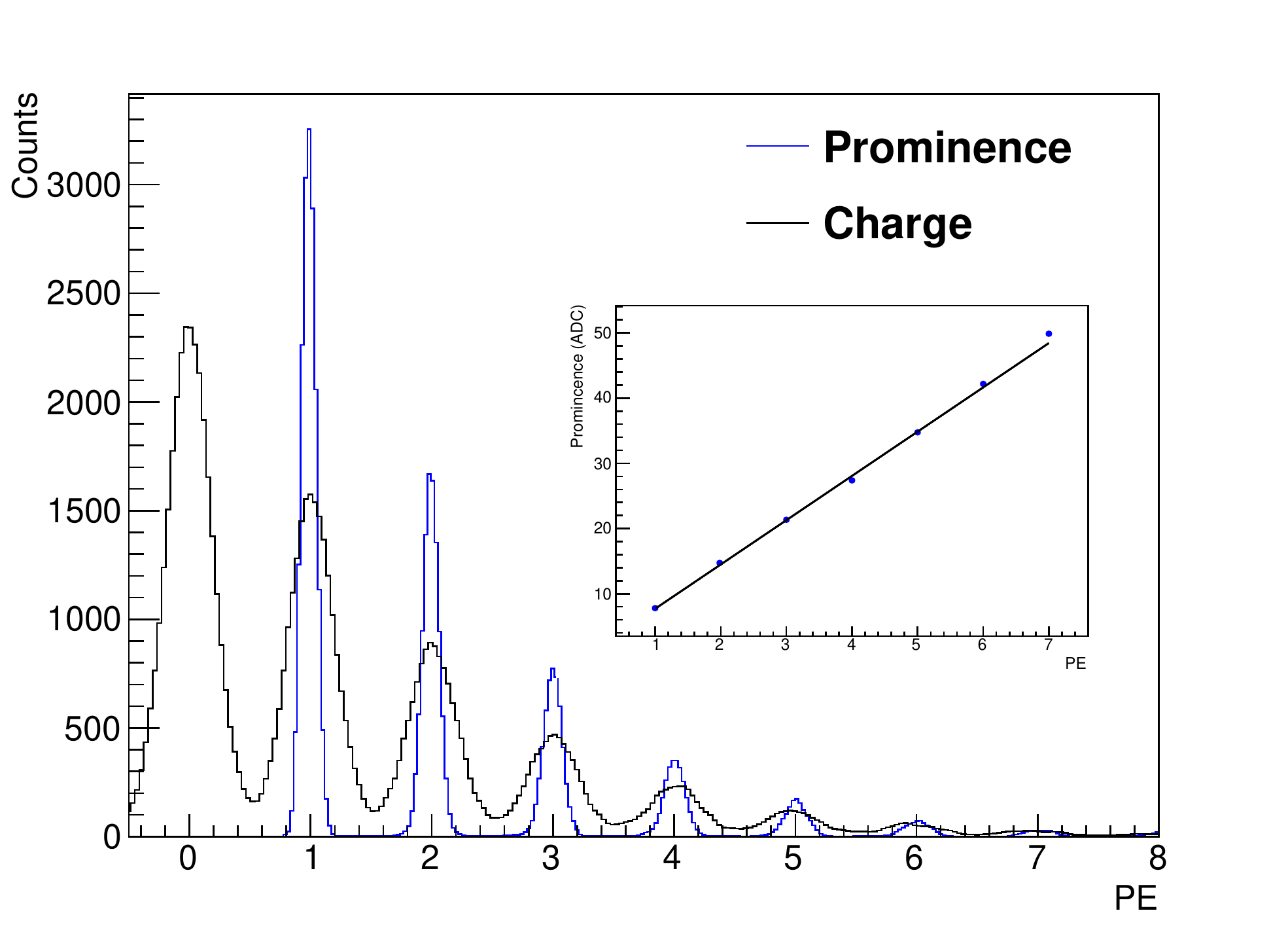}
\caption{Charge and prominence distributions, in number of photoelectrons (PE), obtained from a typical laser calibration run for a top SiPM. The inset shows the linear fit used for the calibration between prominence and number of photoelectrons and for the extraction of the Single photoElectron Response (SER). A similar linear calibration is also performed for the charge distribution.} \label{fig:sers}
\end{figure}

The charge and prominence distributions obtained from a typical laser calibration run for a top SiPM are shown in Fig.\,\ref{fig:sers}. The distributions are shown in number of photoelectrons (PE), following the procedure described below. Note that the prominence method does not produce the pedestal peak at $N_{PE}=0$, which is by contrast visible in the charge distribution.
The individual spectra are fitted with a sum of Gaussian distributions to model the response
to $1 \ldots N$ photoelectrons. A linear fit on the mean value of each peak vs.\ the number of photoelectrons is then performed and the SER is evaluated as the slope of this line, as shown in the inset of Fig.~\ref{fig:sers}. 
%The noise pedestal which is visible in the charge distribution is centered around zero, as noise is averaged out along
%the 4~$\mu$s integration gate. 
The standard deviation $\sigma_1$ of the single-photon peak 
%\footnote{The signal induced by a single photon has an amplitude of approximately 3\,mV. Saturation due to the front-end electronics begins at $\sim$550\,mV, which corresponds to approximately 170 simultaneous photons on a given channel. Since the SiPMs on the bottom tile are read out in groups of six, effects due to saturation are typically first observed on these channels.}
is 0.16 (0.20)\,PE for the charge
distribution of bottom (top) channels; the prominence method gives a significantly better resolution, with
$\sigma_1 = 0.076$ (0.057)\,PE for the bottom (top) tile.
%The signal-to-noise ratio, defined as the ratio between the distance
%between the $N=1$ and $N=2$ peaks and the rms fluctuations of the noise pedestal, namely
%\begin{equation}
%SNR = \frac{\mu_2 - \mu_1}{\sigma_0},
%\end{equation}
%is $SNR = 6.3 (5.1)$ for the charge distribution of bottom(top) SiPM; it cannot be evaluated for the prominence
%distribution due to the lack of the pedestal.

\begin{figure}
\centering
\includegraphics[trim=0 0 1.4cm 0.5cm,clip,width=0.95\columnwidth]{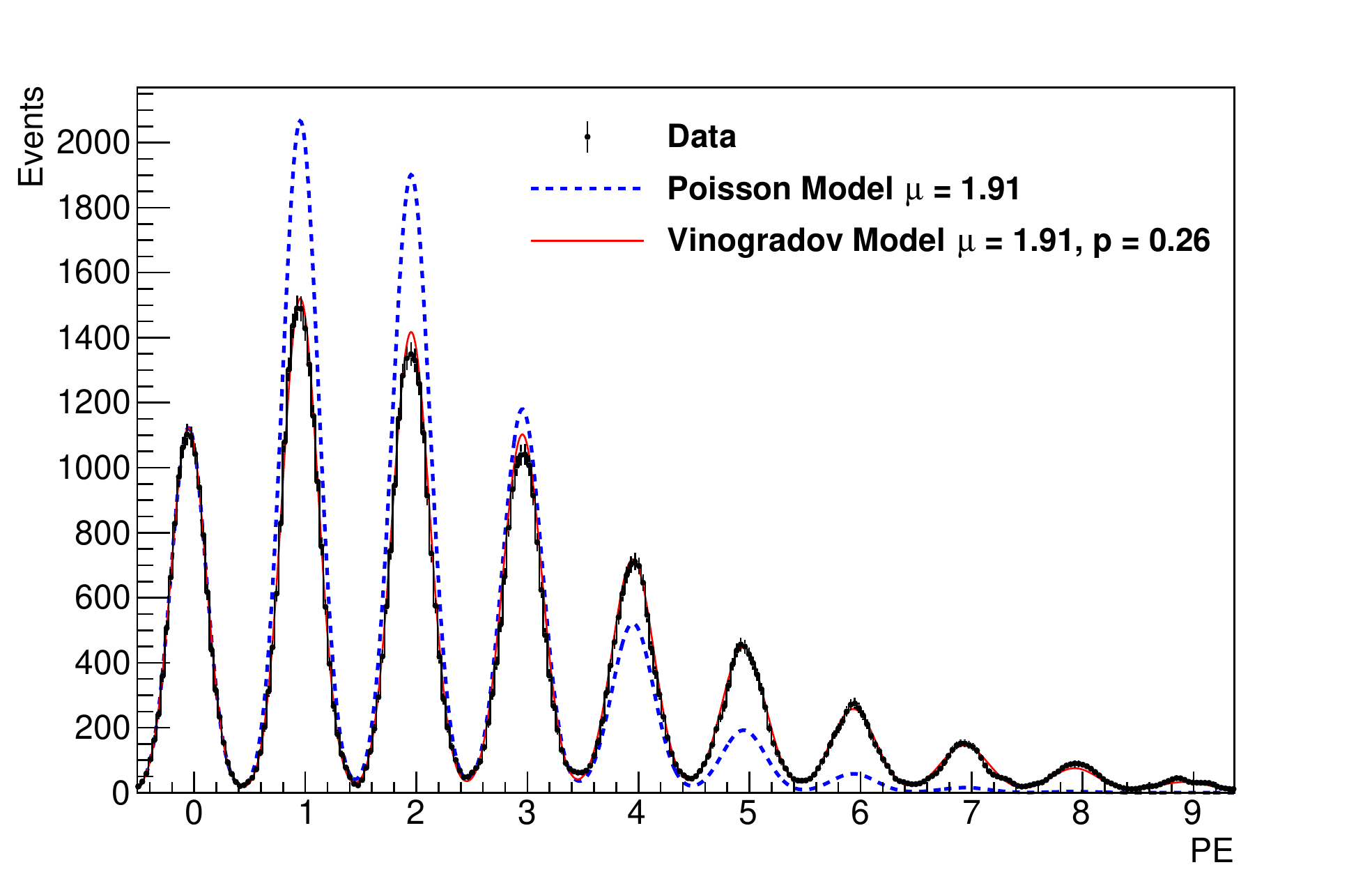}
\caption[Single-photon spectrum with a compound Poisson model]{A typical charge distribution, in number of photoelectrons, for a bottom channel (black histogram), superimposed with a pure Poisson model with $\mu = 1.91$ (blue dashed line) and a compound Poisson model of Eq.\,\ref{eq:vinogradov} with $\mu = 1.91$ and $p = 0.26$ (red solid line).
}
\label{fig:vinogradovfit}
\end{figure}

Due to the effects of afterpulsing and crosstalk, the response of a SiPM to excitation by 
a primary photon corresponds on average to the measurement of more than one photoelectron\footnote{In this work, \textit{photons} are the quanta of energy incident on the SiPM, while \textit{photoelectrons} are the quanta of energy measured by the SiPM.} in the SiPM.
%Each time an SPAD is excited by one photon, its response corresponds, on average, 
%to more than one photon due to the effects of afterpulsing and crosstalk. 
In this respect, the probability to detect $N$ photoelectrons does not follow a Poisson distribution; it can instead be described by the Vinogradov model~\cite{vinogradov2009}, which employs a compound Poisson distribution
\begin{equation}
%f_n(\mu,p) = \frac{e^{-\mu} \sum_{i=0}^{n} B_{i,n} [\mu(1-p)]^i p^{n-i}}{n!}, \label{eq:vinogradov}
f_N(\mu,p) = e^{-\mu} \sum_{i=0}^{N} \frac{B_{i,N}}{N!} [\mu(1-p)]^i p^{N-i}, \label{eq:vinogradov}
%f_N(\mu,p) = \frac{e^{-\mu}}{N!} \sum_{i=0}^{N} B_{i,N} [\mu(1-p)]^i p^{N-i}, \label{eq:vinogradov}
\end{equation}
where $\mu$ is the mean number of primary photoelectrons, $p$ is the probability for a primary photoelectron to trigger 
a secondary emission in the SiPM, and the coefficient $B_{i,N}$ is
 \[ B_{i,N} = \begin{cases} 
      1,  & i=0,~N=0\\
      0, & i=0,~N>0 \\
      \frac{N!(N-1)!}{i!(i-1)!(N-i)!}, & \textrm{otherwise}.
   \end{cases}
\]
Defining the coefficient of duplication $K_{dup} = \frac{p}{1-p}$, the value (1+$K_{dup}$) then represents the total number of photoelectrons detected for each primary photon that induces an excitation
in the SiPM. 
%The coefficient $B_{i,n}$ is 1 if $i=0$ and $n=0$; 0 if $i=0$ and $n>0$; and $\frac{n!(n-1)!}{i!(i-1)!(n-1)!}$ otherwise. 
%The mean ($EX$) and the variance ($Var(X)$) of the compound Poisson distribution are
%\begin{equation}
% EX = \mu \cdot(1+K_{dup})
%\end{equation}
%%
%\begin{equation}
% Var(X) = EX\cdot(1+2K_{dup});
%\label{fano}
%\end{equation}
%here, $\mu$ is the mean of the pure Poisson distribution, and the term 1+2K$_{dup}$ is the so-called 
%Fano factor. 
%The ideal case (and the Poisson distribution) is restored for $p=0$, which yields 
%$K_{dup} = 0$ and $EX = Var(X) = \mu$.
The parameters $\mu$ and $p$ of the Vinogradov model of Eq.\,\ref{eq:vinogradov} are calculated by running 
a maximum likelihood fit on the amplitude distribution of the $0, 1 \ldots N$ photoelectron 
peaks from Fig.\,\ref{fig:sers}. The output of the fit is shown in Fig.\,\ref{fig:vinogradovfit}: data from a
bottom channel are superimposed with a Poisson distribution ($\mu=1.91$, $p=0$) and a compound Poisson distribution ($\mu=1.91$ and $p=0.26$).
\color{black}
 Typical values of $K_{dup}$ obtained for the individual channels range between 0.31 and 0.37, with statistical uncertainties from the fit of approximately 3\%. The effect due 
to afterpulsing beyond the 4\,\textmu s integration window is evaluated to be well below the statistical 
uncertainty.
%

%Time stability
The relative fluctuation of the SER values, calculated from all 42 available laser calibrations,  
is 0.7\% (1.0\%) rms for bottom (top) channels. This is consistent with the level of temporal variation in the output voltage of the power supply. Variations of the SER between consecutive laser calibrations are well below 2\% for all channels, except for two SiPMs of the top tile that occasionally exhibited variations up to 6--7\%. The relative fluctuation on $K_{dup}$ over time are 3.0\% (3.6\%) rms for bottom (top) channels and are the same order of magnitude as the typical statistical uncertainties of the individual fits.

%Bias correction
The SiPM bias circuit contains series resistors that cause an effective reduction of the overvoltage applied to the SiPMs when the leakage current of the devices is high. This typically occurs when the SiPMs are exposed to a significant amount of light, e.g.\ due to a high interaction rate from intense radioactive sources or to large S2 signals from high multiplication in the gas. Currents recorded by the slow control system range from $0.5$\,\textmu A  at null
field (in absence of S2) up to 11 (18)\,\textmu A in specific high-field configurations for the bottom (top) tile. The drop in overvoltage causes a reduction of the SiPM gain and SER, which must be properly taken into account. The correction is approximately 0.5\%/\textmu A, derived from studies of isolated photoelectrons and the dark count rate, and including also the known variation of the gain and photon detection efficiency as a function of overvoltage. At the highest field value reported here, the maximum applied correction to the overall TPC light response is below 5\%.
% From Marco: The correction is about 0.5 %/μA, to which we assign conservatively a 50% uncertainty; this correction is derived by studying the SER from dark rate or isolated photoelectrons in the tail of physics events as a function of the measured bias current in runs acquired with the TPC exposed to sources providing different event rates and by also considering the known variation of the photon detection efficiency as a function of the applied overvoltage. “
%

\section{Scintillation (S1) response} \label{sec:s1}
Measurements of the \TPC\ response were performed by irradiating the detector with an
external \Am\ source, which emits monoenergetic 59.54\,keV $\gamma$-rays. The scintillation response is studied by operating the \TPC\ in 
single-phase mode, i.e.\ filled with liquid only, and at null field ($\edrift=\eex=\eel=0$). For each trigger, signals from the 28 SiPM readout channels are acquired for a total of 20\,\textmu s, 30\% of which contains pre-trigger data used for a precise calculation of the baseline. Individual traces are then baseline subtracted, corrected for the different gains of the SiPMs, and summed. The summed waveform is scanned by a pulse-finder algorithm to search for signals. Scintillation signals from electron recoils larger than $\sim$20\,PE are efficiently identified by the algorithm. Once a candidate signal is found, its associated time is taken as that corresponding to the constant fraction value of 70\% of the maximum. The S1 signal is evaluated by using the total charge measured by the SiPM: the voltage signals are integrated in a 12\,\textmu s window starting 3\,\textmu s before the associated time; the longer integration gate with respect to laser calibrations is necessary to account for the scintillation light emitted by the argon triplet state, which has a time constant of approximately 1.6\,\textmu s. 

\begin{figure}
\centering
\includegraphics[trim=0 0 1.5cm 0.5cm,clip,width=0.95\columnwidth]{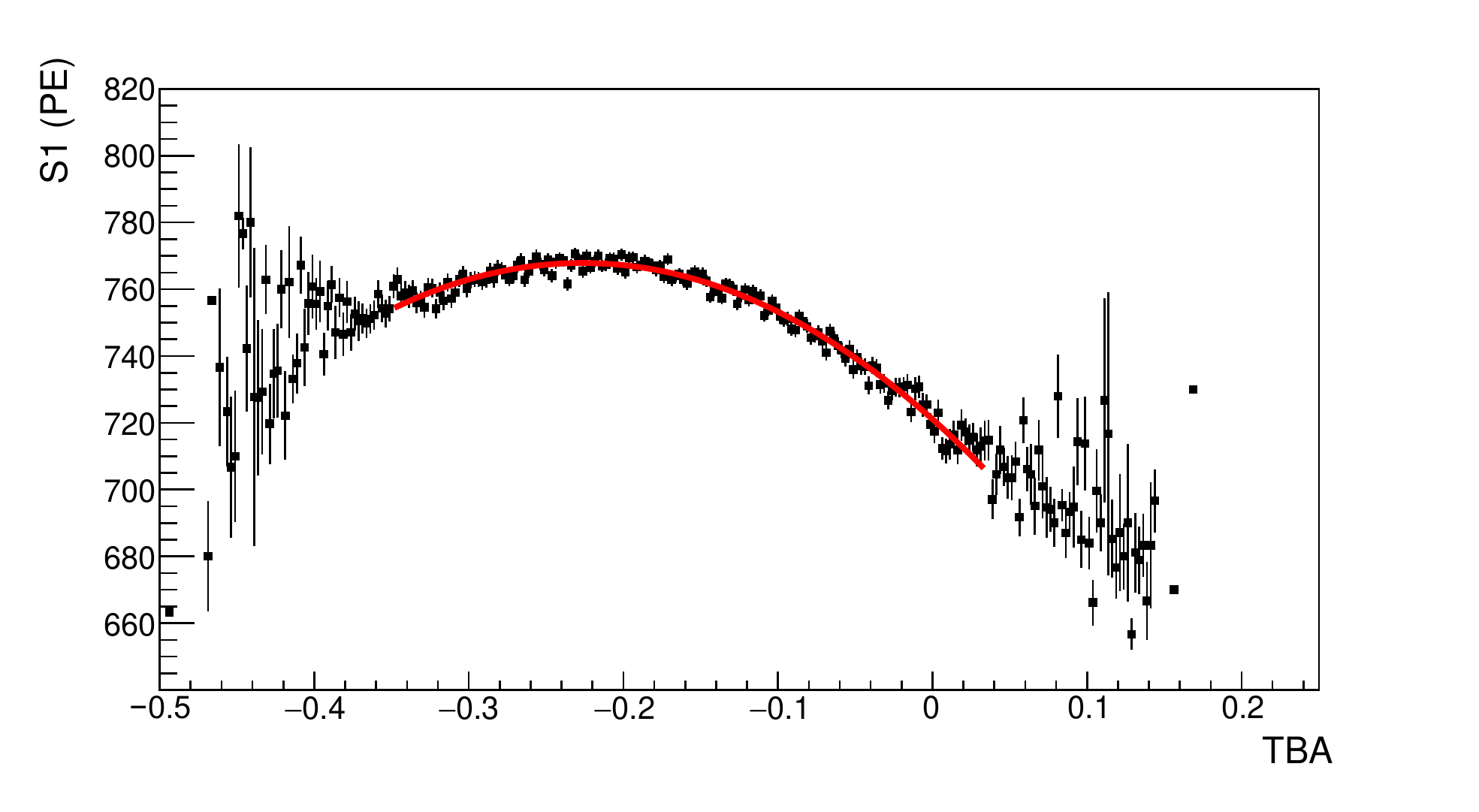}
\caption[S1 TBA]{S1 top-bottom asymmetry (TBA) before the TBA correction is applied, for full-energy \Am\ events taken in single-phase mode at null field. }
\label{s1-tba}
\end{figure}

As previously observed in DarkSide-50~\cite{Agnes:2017cl}, the distribution of photons on the top and bottom tiles depends on the position of the primary scintillation event along the drift ($z$) axis. The top-bottom asymmetry (TBA) is defined as 
\begin{equation}
\label{eq:tba}
TBA = \frac{S1_{top}-S1_{bottom}}{S1_{top}+S1_{bottom}} 
\end{equation}
where $S1_{top}$ and $S1_{bottom}$ are the S1 signals globally detected by the top and bottom channels, respectively. 
The TBA is thus a measure of the asymmetry in the S1 signal collected by the top and bottom tiles and is therefore correlated 
with the $z$ position of the event.
Fig.\,\ref{s1-tba} shows the S1 scintillation signal as a function of TBA for an \Am\ calibration run taken in single-phase mode at null field. Events are required to have only one scintillation signal in the entire acquisition window and an S1 signal compatible with the full absorption of the 
59.54\,keV photon from \Am.
As can be seen in Fig.\,\ref{s1-tba}, the size of the S1 signal varies with TBA. To correct for this effect, the distribution is fitted with a second-order polynomial and the measured S1 signals are scaled event-by-event according to the fit, resulting in a flat distribution. % shown in the bottom panel of Fig.\,\ref{s1-tba}.
\begin{figure}
\centering
\includegraphics[trim=0 0 1.0cm 0.5cm,clip,width=0.95\columnwidth]{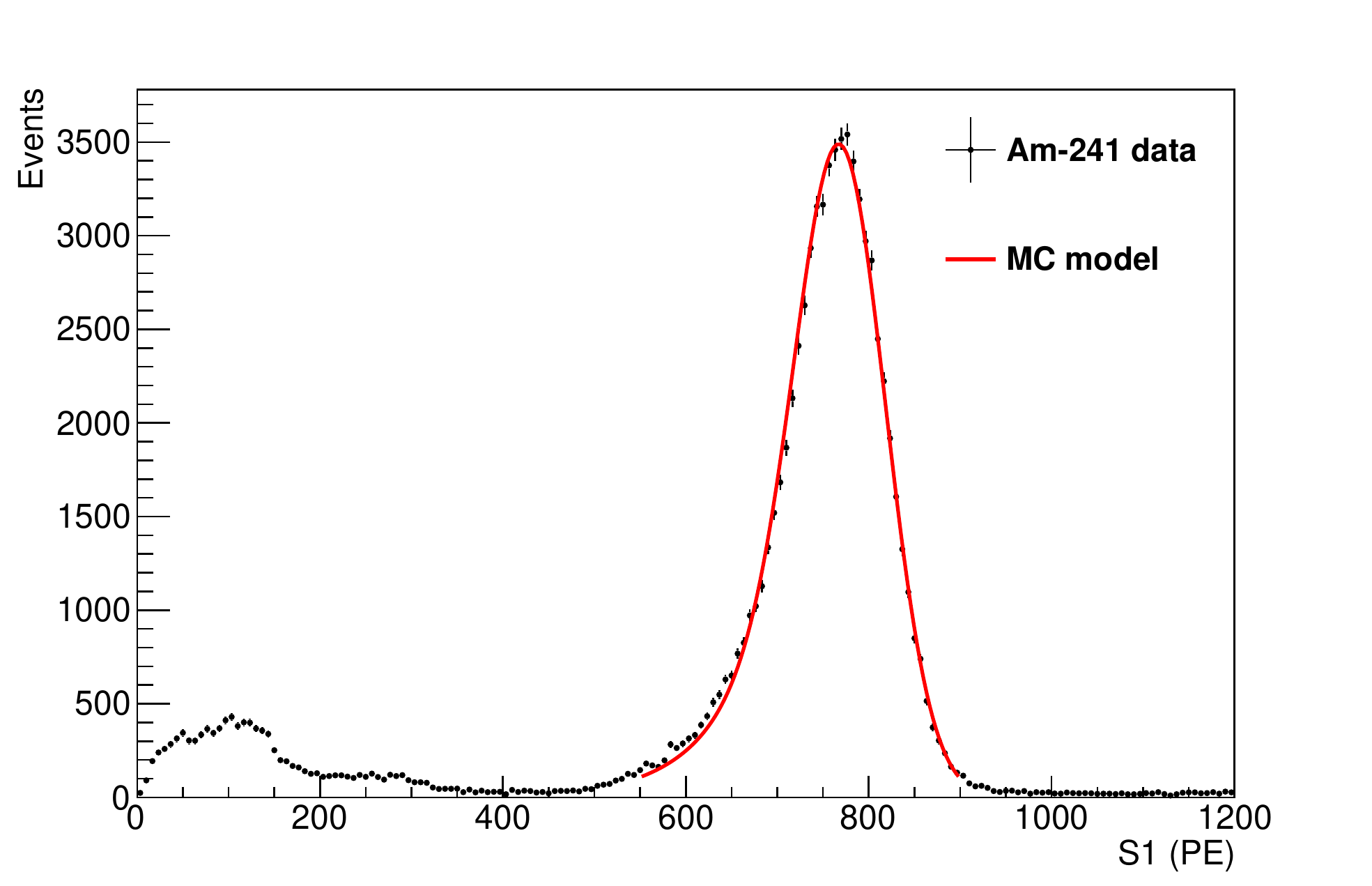}
\caption[S1 \Am\ spectra]{S1 distribution for a \Am\ measurement taken in single-phase mode at null field, with TBA correction applied. The best fit to the MC model discussed in the text is superimposed.} \label{fig:s1-am-fit}
\end{figure}

In general, the light yield ($Y$) is calculated as the number of photoelectrons detected per unit of deposited energy. To account for secondary emissions due to crosstalk and afterpulsing (see Sec.\,\ref{sec:ser}), the corrected light yield is calculated from the raw, measured light yield according to
\begin{equation}
Y_{corr} = \frac{Y_{raw}}{1+\langle K_{dup} \rangle } \label{correctedS1}, 
\end{equation}
where  $\langle K_{dup} \rangle$ is the channel-averaged duplication factor. 

The S1 light yield ($Y_1$) is calculated using the full-energy \Am\ peak in the S1 spectrum, corresponding to 59.54\,keV. The S1 distribution following the TBA correction is fitted to a template computed as the numerical convolution of the \Am\ spectrum from MC, which describes the true energy deposited in the \TPC, with a Gaussian smearing function to account for the detector resolution. The free parameters of the fit are the total light yield ($\mu$) and the standard deviation ($\sigma$) of the smearing function.
%The 
%profile of energy resolution vs. energy for the gaussian smearing is calculated as 
%\begin{equation}
%\sigma = \sqrt{\sigma_0^2 + F \cdot N_{PE}} = \sqrt{\sigma_0^2 + F \cdot E \cdot LY}, 
%\end{equation}
%with the term $F$ (``Fano factor'') describing the degradation with respect to the pure statistical 
%fluctuations. The light yield $LY$, the Fano factor $F$ and $\sigma_0$ are free parameters in 
%the fit. According to the Vinogradov model, it is expected to be $F = 1 + 2 K_{dup} \sim 1.6$.
Data from a \Am\ calibration run taken in single-phase mode at null field are shown 
in Fig.\,\ref{fig:s1-am-fit}, along with the best fit to the smeared MC template:
the uncorrected light yield and energy resolution are $Y_{1,raw} = (13.03 \pm 0.05)$\,PE/keV and $\sigma/\mu = 6.4\%$, respectively.
%the resolution corresponds to $F = 3.16$. 
Applying the TBA correction improves the resolution from $\sigma/\mu = 6.6\%$ to 6.4\%. 
The corrected light yield is calculated from Eq.\,\ref{correctedS1} as $Y_{1,corr}=(9.80 \pm 0.13)$\,PE/keV at null field. The uncertainty on $Y_{1,corr}$ is derived from the combination of statistical and systematic uncertainties on $Y_{1,raw}$ and $\langle K_{dup} \rangle$. For the remainder of this work, S1 signal strengths and light yields are always quoted following the TBA correction, unless otherwise noted.
%Time stability
The time stability of the S1 response of the \ReD\ \TPC\ in single-phase operation is verified using four \Am\ calibrations taken in equivalent conditions throughout the operational period. The position of the full-energy \Am\ peak is found to be reproducible to within 2\%.

%Prominence
Improved performance in regards to energy resolution can be achieved using the prominence approach, 
which allows for the identification of single photoelectrons. 
%This technique is particularly effective in the regime of interest for DarkSide-20k, which will employ 8,000 channels. 
However, since the occupancy of each channel in \ReD, defined as the number of photoelectrons per channel, is significantly greater than 1 due to the limited size of the chamber and the low number of readout channels, the pile-up of photoelectrons causes a strong reconstruction inefficiency resulting in a non-linear response. 
%For the \Am\ peak considered here, the light yield obtained using the prominence method corresponds to $\sim$2/3 of that obtained from charge integration, albeit with better resolution ($\sigma/\mu = 5.6\%$ after TBA correction), a promising achievement in view of DarkSide-20k. 
In this work, S1 and S2 signals are thus calculated using charge integration. 

\begin{figure*}
\centering
\includegraphics[trim=0 0 1.2cm 0.5cm,clip,width=0.95\columnwidth]{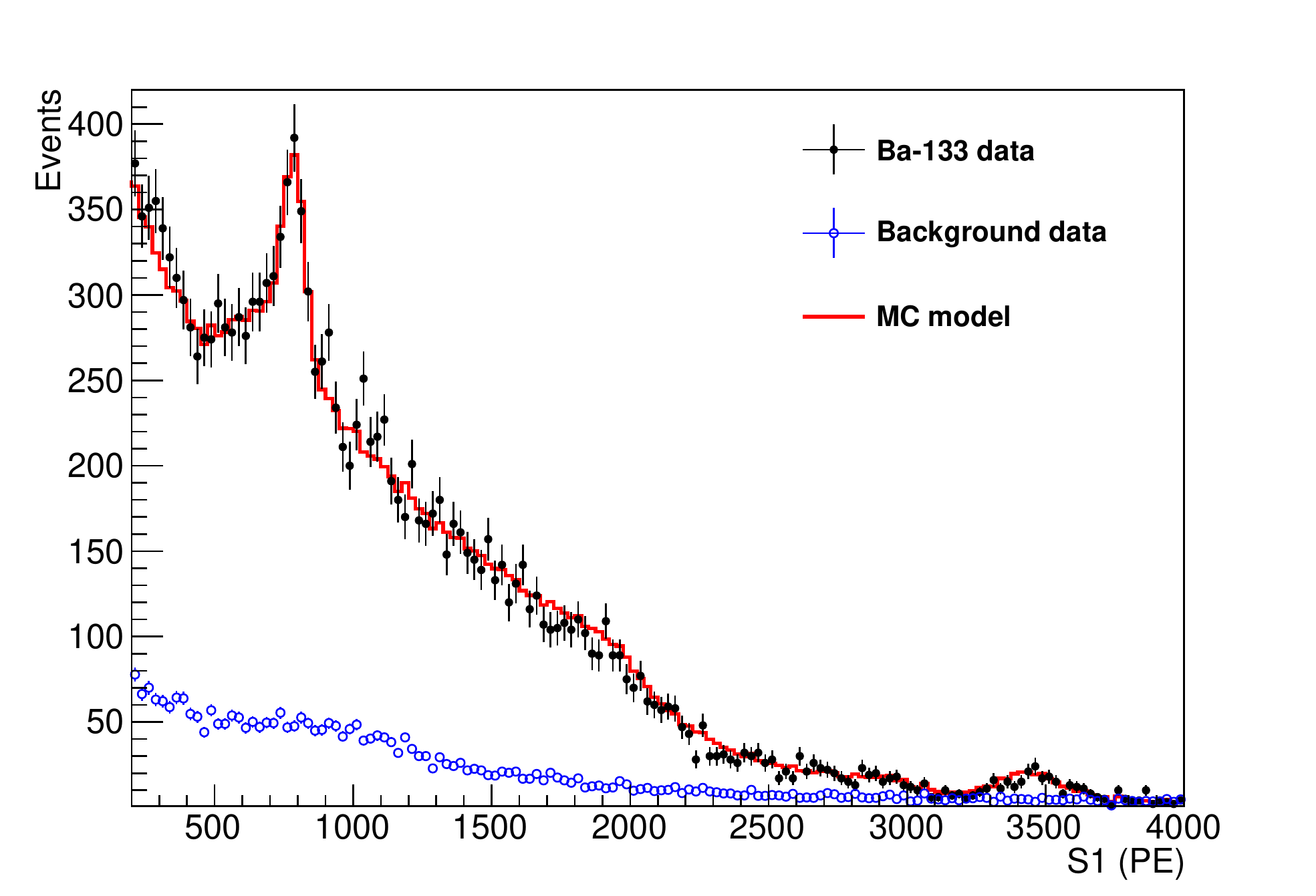}
\hspace{0.2cm}
\includegraphics[trim=0 0 1.2cm 0.5cm,clip,width=0.95\columnwidth]{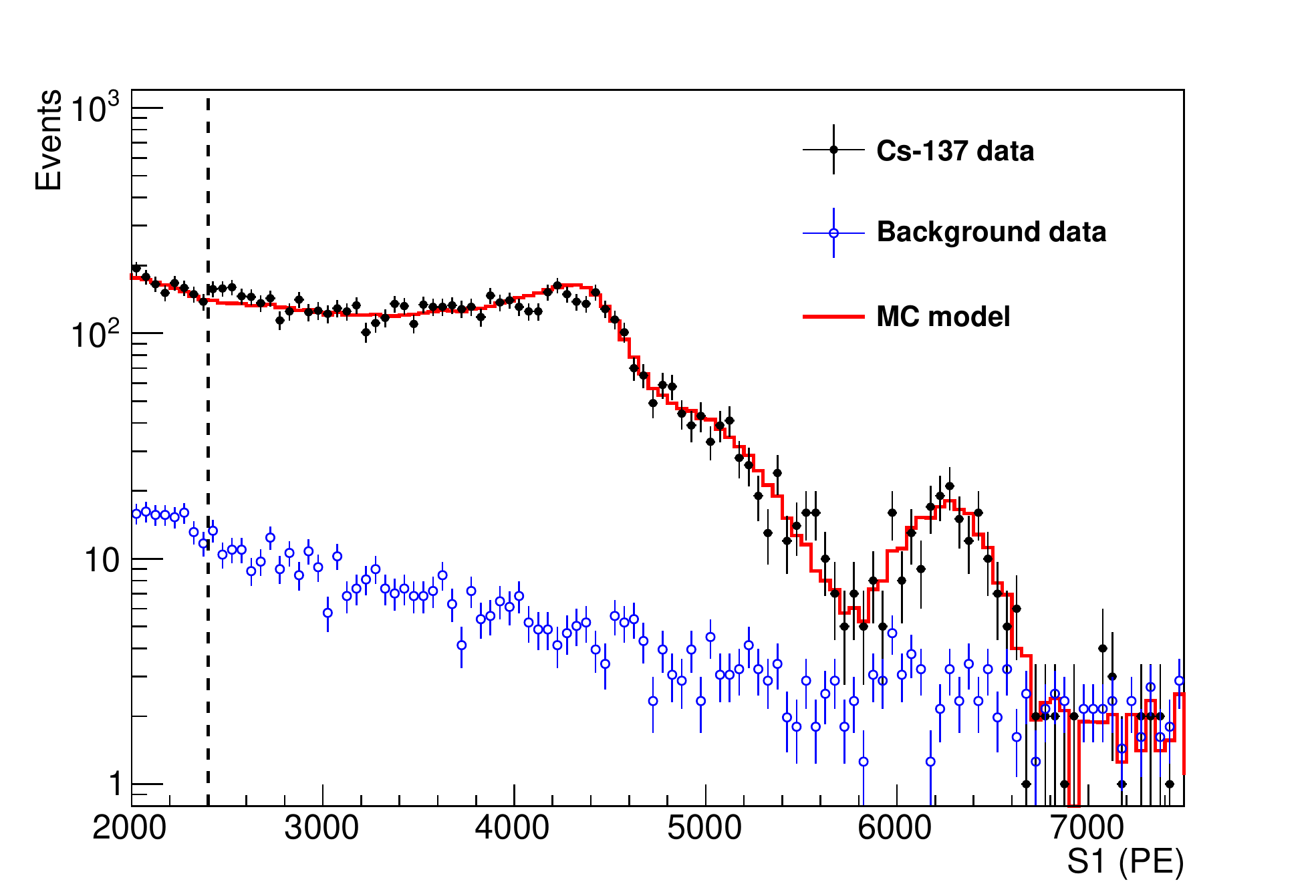}
\caption[S1 spectra from $^{133}$Ba and $^{137}$Cs at null field]{S1 distribution
from $^{133}$Ba (\textit{left}) and $^{137}$Cs (\textit{right}) calibrations taken in single-phase mode at null field, with no TBA correction applied. Each distribution is superimposed with the normalised environmental background and the best fit to a template (Monte Carlo signal + background data) that accounts for detector resolution. The dotted line at 2400 PE marks the start of the fit window for the $^{137}$Cs S1 spectrum.} \label{fig:michaelnullfield}
\end{figure*}

%Other sources
Measurements of the scintillation light response at null field were also taken with other external sources, including $^{133}$Ba and $^{137}$Cs. Similar to that for \Am, the S1 light yield is calculated by fitting the measured S1 distribution to an MC template convoluted with a Gaussian distribution representing the detector resolution. For each source, the Compton spectrum and the photopeak (81\,keV for $^{133}$Ba and 662\,keV for $^{137}$Cs) are fitted with independent light yield and resolution parameters to account for their energy dependence. Fig.\,\ref{fig:michaelnullfield} shows the S1 distribution and the best fit for calibration runs taken with $^{133}$Ba and $^{137}$Cs in single-phase mode at null field.
The measured light yields are $Y_{1,corr}=(9.70 \pm 0.14)$\,PE/keV at 81\,keV and $Y_{1,corr}=(9.48 \pm 0.12)$\,PE/keV at 662\,keV. No TBA correction has been applied to these spectra since the resulting impact on the light yield at the photopeak is negligible ($<0.4\%$).
%No TBA correction is applied on the $^{133}$Ba ($^{137}$Cs) spectra, but a
%0.8\% (0.9\%) systematic uncertainty is included on the light yields above, estimated from the
%variation of S1 over TBA.
%
% Michael's email
% For the null-field Ba photopeak: LY = 9.70$^{+0.05}_{-0.02}$ PE/keV. Syst+stat: 1.3%
% For the null-field Cs photopeak: LY = 9.48 $\pm$ 0.05 PE/keV Syst+stat: 1.3%
%
%

%Dual-phase runs
Dedicated measurements with the \Am\ source were also taken in double-phase mode (i.e.\ in the presence of a gas pocket) at null field. Variations at the level of only 1\% were found between the measured light yields in single-phase and double-phase operation at null field, compatible with expected fluctuations over time. %This indicates that the gas pocket, despite its different refractive index relative to \LAr, does not substantially affect the light collection efficiency within the active volume. 

\section{Scintillation-ionisation (S1-S2) response} \label{sec:s2}
\label{sec:chargeresp}
Detection of the ionisation signal (S2) requires drifting the free electrons from the interaction point to the liquid-gas interface, extracting them from the liquid to the gas, and accelerating them in the gas to produce electroluminescent light. The TPC must therefore be operated in double-phase mode, namely with a gas pocket above the liquid phase, and with the appropriate electric field in the three regions of the \TPC: $\edrift$, $\eex$ and $\eel$. Since the S2 signal is delayed by several tens of \textmu s with respect to S1 due to the electron drift time, signals from the SiPMs are acquired for a total window of 100\,\textmu s, with approximately 10\% of this time reserved for the pre-trigger. 

%Special runs, meant to study echo S3 events 
%were taken with windows of 80000 or 100000 samples (160 and 200~$\mu$s, respectively). 
%The efficiency of the pulse-finding algorithm for S2 signals cannot be evaluated by using the 
%same method of Fig.~\ref{fig:reco-eff}, as the DAQ trigger is typically on S1. However, since the algorithm 
%applies to the waveform a moving average of $2~\mu$s width, S1 signals are substantially smeared 
%out in time and look hence fairly similar to S2's. This arguments for a S2 efficiency of the 
%pulse-finder which is comparable to what shown in Fig.~\ref{fig:reco-eff} for S1.

\subsection{Drift time distribution and drift velocity} \label{sec:drifttime}
\begin{figure}
\centering
\includegraphics[trim=0 0 1.4cm 0.5cm,clip,width=0.95\columnwidth]{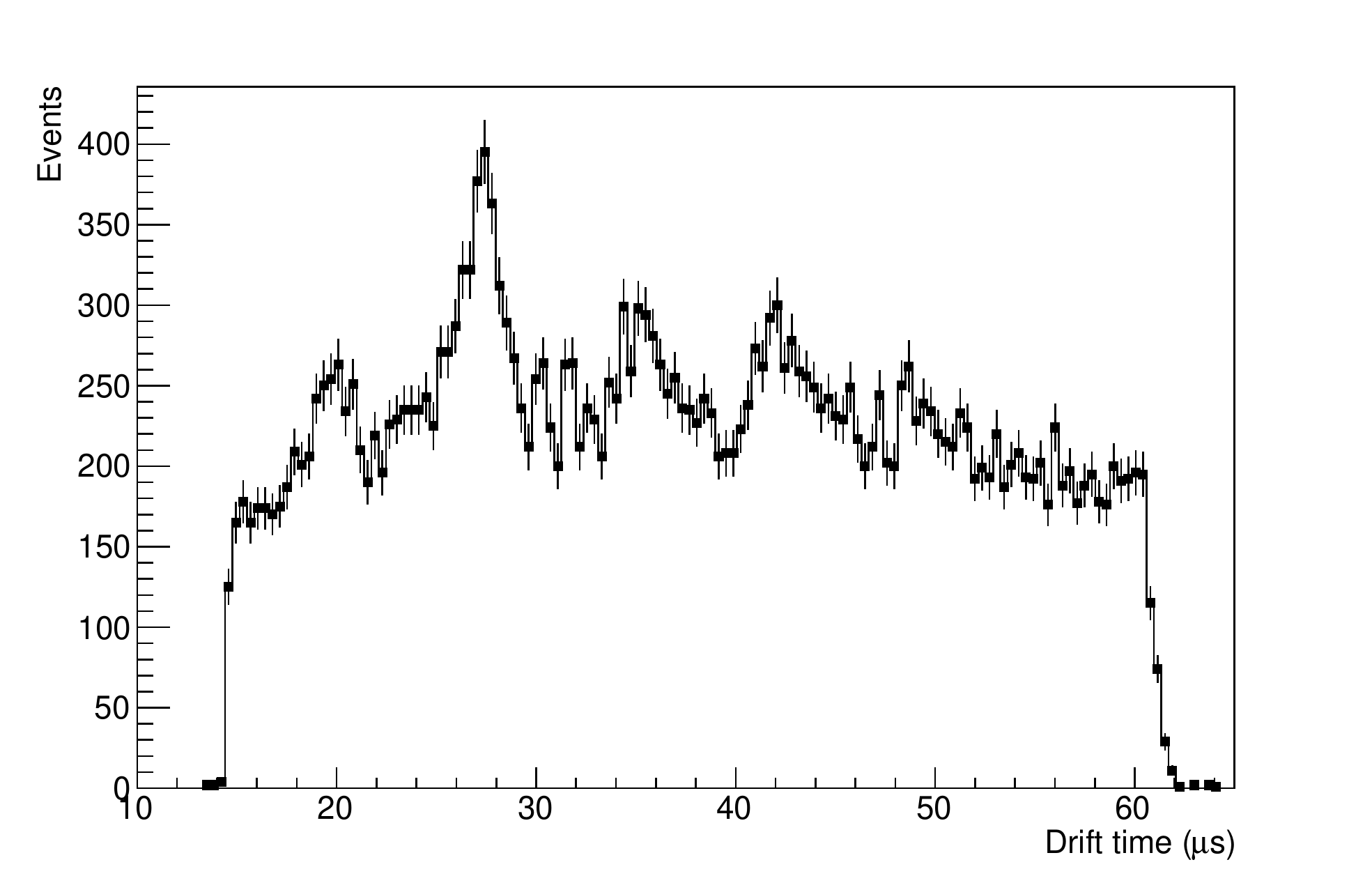}
\caption[Drift time for a \Am\ calibration taken at $\langle \edrift \rangle = 183$\,V/cm drift field]{Drift time distribution for an \Am\ calibration run taken at $\edrift =183$\,V/cm. The valleys in the distribution are due to the presence of the copper field-shaping rings and are reproduced by MC simulation.}
\label{fig:tdrift}
\end{figure}

The total drift time between the onset of the prompt scintillation signal (S1) and the delayed ionisation signal (S2) gives information about the $z$ coordinate of the primary interaction. Fig.\,\ref{fig:tdrift} shows the drift time distribution for data taken with an external \Am\ source at $\edrift =183$\,V/cm. The distribution features a cutoff at $\sim$62\,\textmu s, corresponding to the time needed for an electron produced at the cathode to travel upwards to the electroluminescence region. The turn-on at 12\,\textmu s is due to the efficiency of the reconstruction, which is unable to fully resolve the separation between S1 and S2 signals below this time, and to the selection cuts intended to remove pile-up events. The valleys in the drift time distribution of Fig.\,\ref{fig:tdrift} are due to the presence of the copper field-shaping rings, which absorb a portion of the $\gamma$-rays from the external \Am\ source. This behaviour is reproduced by the MC simulation.

Measurements of the drift time can be used to determine the electron drift velocity in \LAr\ as a function of $\edrift$ and the electron mobility at the operating temperature. The temperature of the \LAr\ here is  $T=(88.17 \pm 0.05)$\,K, calculated according to the saturation PT curve of \LAr, evaluated at the average measured value of the atmospheric pressure and corrected for the additional contribution from the hydrostatic pressure of the \LAr\ column above the \TPC.
Five measurements were taken at $\edrift$ between 75 and 1000\,V/cm, with $\eex$ and $\eel$ set to their standard reference values of 3.8\,kV/cm and 5.7\,kV/cm, respectively. The electron drift velocity in \LAr\ is calculated using the distance between the cathode and grid, $D=(49.1 \pm 0.5)$\,mm, taking into account the thermal contraction of PTFE at 88\,K~\cite{Kirby:1956}. The time for electrons to drift the distance $D$ can be estimated as $(T_{max} - T_{min} + t_0)$, where $T_{max}$ is the cutoff of the drift time distribution, $T_{min}$ is the time needed for the electrons to drift across the liquid layer between the grid and the gas interface, and $t_0$ accounts for the diffusion along the drift direction, which causes the initial electrons to arrive earlier than the centroid of the ionisation cloud. The drift time cutoff ($T_{max}$) varies between $\sim$125\,\textmu s 
at $\edrift=75$\,V/cm and $\sim$25\,\textmu s at 1000\,V/cm. Considering an electron velocity of $(3.5 \pm 0.1)$\,mm/\textmu s in liquid argon at $\eex = (3.8 \pm 0.2)$\,kV/cm, the transit time through the 
extraction region, namely the $(3 \pm 1)$-mm layer of \LAr\ above the grid, is $T_{min}=(0.9 \pm 0.3)$\, \textmu s.
The diffusion correction ($t_0$) is calculated according to the analytical parametrization from \cite{Agnes:2018hvf}; $t_0$ is on the order of a few \textmu s for small $\edrift$, becoming negligible ($< 0.01$\,\textmu s) above 600\,V/cm. 

\begin{figure}
\centering
\includegraphics[trim=0 0 1.4cm 0.5cm,clip,width=0.95\columnwidth]{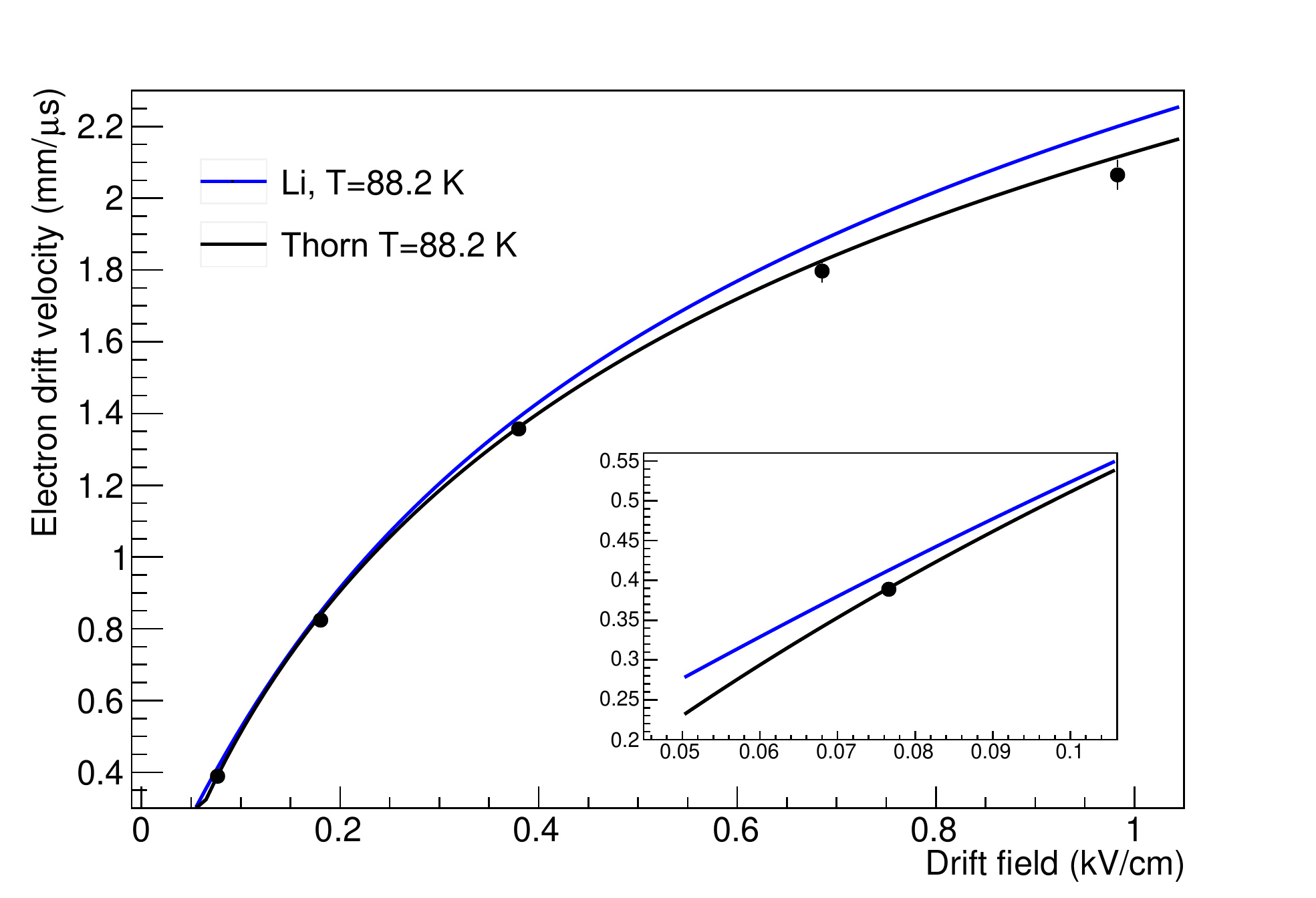}
\caption[Drift velocity in \LAr\ vs.\ electric field, compared to data]{Electron drift velocity in \LAr\ as a function of the drift field ($\edrift$), calculated according to the parameterisations in \cite{ThornLArProperties} (black line) and \cite{Li:2016dz} (blue line) at $T= 88.2$\,K, superimposed with the experimental data. The inset shows a zoomed view of the data point at $\edrift = 76.7$\,V/cm.}\label{fig:driftvelocity}
\end{figure}

The measured electron drift velocity in \LAr\ is shown in Fig.\,\ref{fig:driftvelocity}, in comparison to two parameterisations from the literature~\cite{ThornLArProperties,Li:2016dz} calculated at a \LAr\ temperature of $88.2$\,K. %The parameterisation from \cite{ThornLArProperties} uses the functional 
%form from \cite{Kalinin:1996} to fit experimental data from \cite{Amoruso:2004ti,Aprile:1985xz}. 
Although the parameterisation from \cite{ThornLArProperties} was developed 
for drift fields above 300\,V/cm, it is in good agreement with the \ReD\ data at low fields ($\chi^2/n_{\rm dof}= 4.7/5$), performing considerably better than that of \cite{Li:2016dz}. The drift velocity distribution was also fitted to the parameterisation of \cite{ThornLArProperties} leaving the temperature as a free parameter, resulting in $T_{fit} = (88.9 \pm 0.4)$~K, compatible with the temperature calculated above.

\subsection{Electron lifetime} \label{sec:elifetime}
\begin{figure}
\centering
\includegraphics[trim=0.2cm 0 1.4cm 0.5cm,clip,width=0.95\columnwidth]{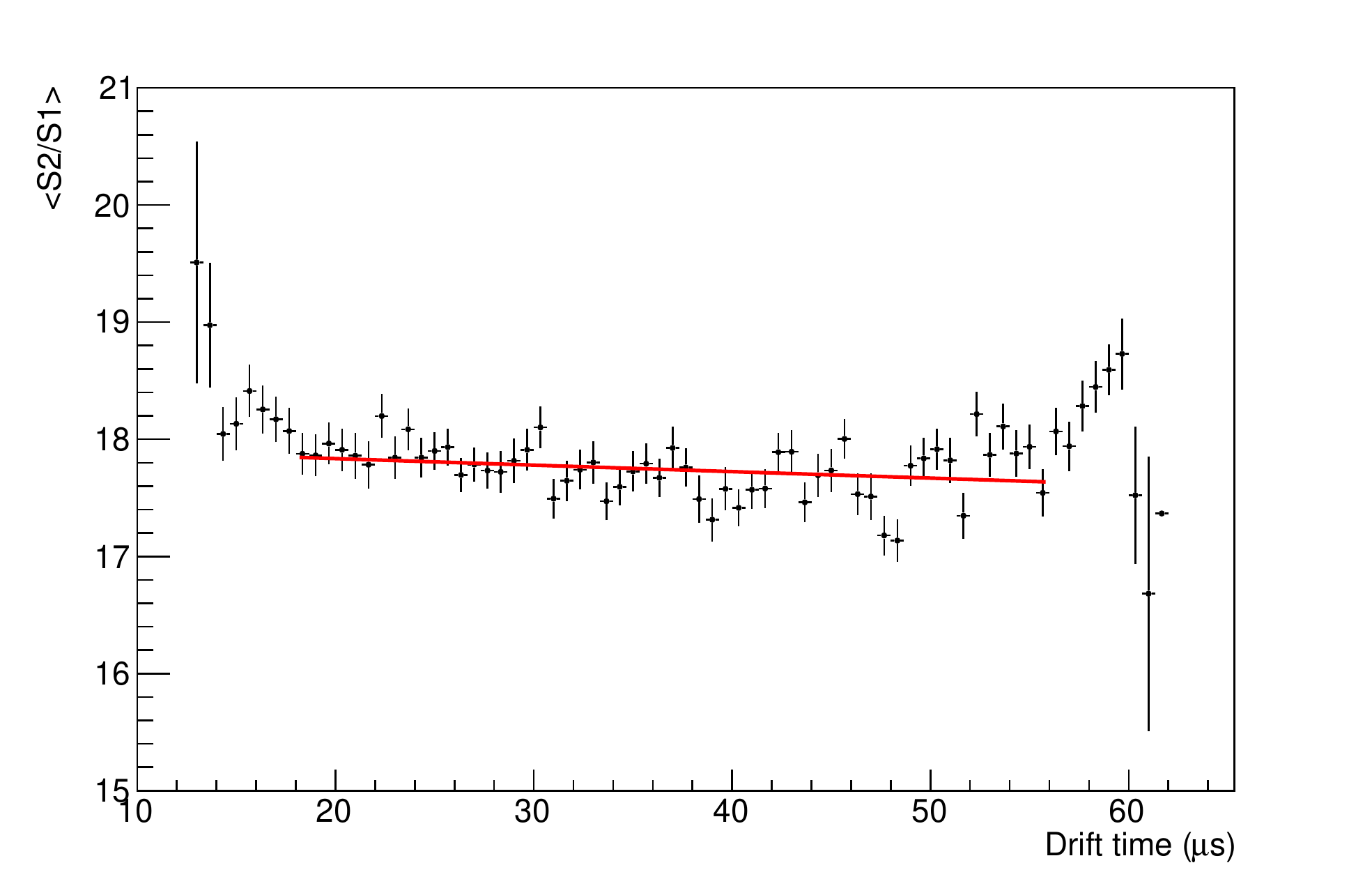}
\caption[$\langle$S2/S1$\rangle$ vs.\ drift time for \Am\ ] 
        {$\langle$S2/S1$\rangle$ vs.\ drift time for an \Am\ calibration run with $\edrift = 183$\,V/cm, together with the best fit according to the parameterisation given in Eq.\,\ref{eq:drifttime}.}        
\label{fig:arpurity}
\end{figure}

Electrons drifting in \LAr\ can be captured by electronegative contaminants such as oxygen and 
nitrogen. The number of free electrons along the drift path ($N_e$) decreases exponentially relative 
to the initial number ($N_0$) as
\begin{equation}
N_e(t) = N_0 e^{-t/\tau} \label{eq:drifttime},
\end{equation}
where $\tau$ is a characteristic time determined by the \LAr\ purity. Under normal operating 
conditions, this lifetime should satisfy $\tau \gg T_{max}$ so that the majority of electrons survive over the entire drift length and arrive to the multiplication region. For this reason, \ReD\ operates a gas recirculation system that continuously purifies the \LAr\ in the \TPC\ (see Sec.\,\ref{sec:cryo}). The electron lifetime is evaluated by measuring the dependence of $\langle$S2/S1$\rangle$ on the drift time. Due to the absorption of electrons, events with a longer drift time (i.e.\ generated closer to the cathode) are expected to have the same S1 signal, but a smaller S2 signal, than events with a shorter drift time. The loss of S2 strength vs.\ drift time ($t_d$) is corrected event-by-event as
\begin{equation}
\textrm{S2}_{corr} = \frac{\textrm{S2}}{e^{-t_d/\tau}}. \label{eq:puritycorrection}
\end{equation}
The distribution of  $\langle$S2/S1$\rangle$ as a function of drift time is shown in Fig.\,\ref{fig:arpurity} for an \Am\ calibration run taken at $\edrift = 183$\,V/cm, following 37 days of recirculation. The red solid line is the best fit to the exponential model of Eq.\,\ref{eq:drifttime}, resulting in an electron lifetime of $\tau = (1.8 \pm 0.6_{\rm \,stat+sys})$\,ms, significantly longer than the maximum drift time in the \TPC. A lifetime greater than 1\,ms is typically calculated for data taken at least two weeks after the initial cool-down, rendering the correction of Eq.\,\ref{eq:puritycorrection} at the level of a few percent. For the remainder of this work, S2 signal strengths are always quoted following this correction and denoted simply as ``S2''. 

\subsection{S2/S1 and ER vs.\ NR discrimination} 
\label{sec:s2s1_er}
The relative ionisation to scintillation yield, or S2/S1 ratio, is a key observable for characterising the performance of a \LAr\ \TPC\ since it provides a handle for discriminating between nuclear recoils (NR) and electronic recoils (ER). Moreover, in view of the physics goals of \ReD, achieving an excellent detector resolution on S2/S1 is essential for precise studies of recombination as a function of recoil angle relative to the $\edrift$ axis.
%In addition, considering the physics goals of \ReD, it is essential to achieve the best possible detector resolution on this measurement in order to accurately study e.g.\ recombination as a function of the recoil angle with respect to the $\edrift$ axis. 
%Moreover, from the instrumental point of view, it encodes the
%capability of  amplification in the gas phase, while its width is sensitive to detector response non uniformities as a
%function of actual position of the events inside the active volume. 
Here the correlations between S1, S2, and S2/S1 are studied using sources of both electron recoils 
(external \Am\ and internal \Kr\ sources) and nuclear recoils. The electric field in the \TPC\ is always kept at the 
reference values: $\edrift =183$\,V/cm, $\eex = 3.8$\,kV/cm and $\eel = 5.7$\,kV/cm.

\begin{figure}
\centering
\includegraphics[trim=0 0 1.2cm 0.5cm,clip,width=0.95\columnwidth]{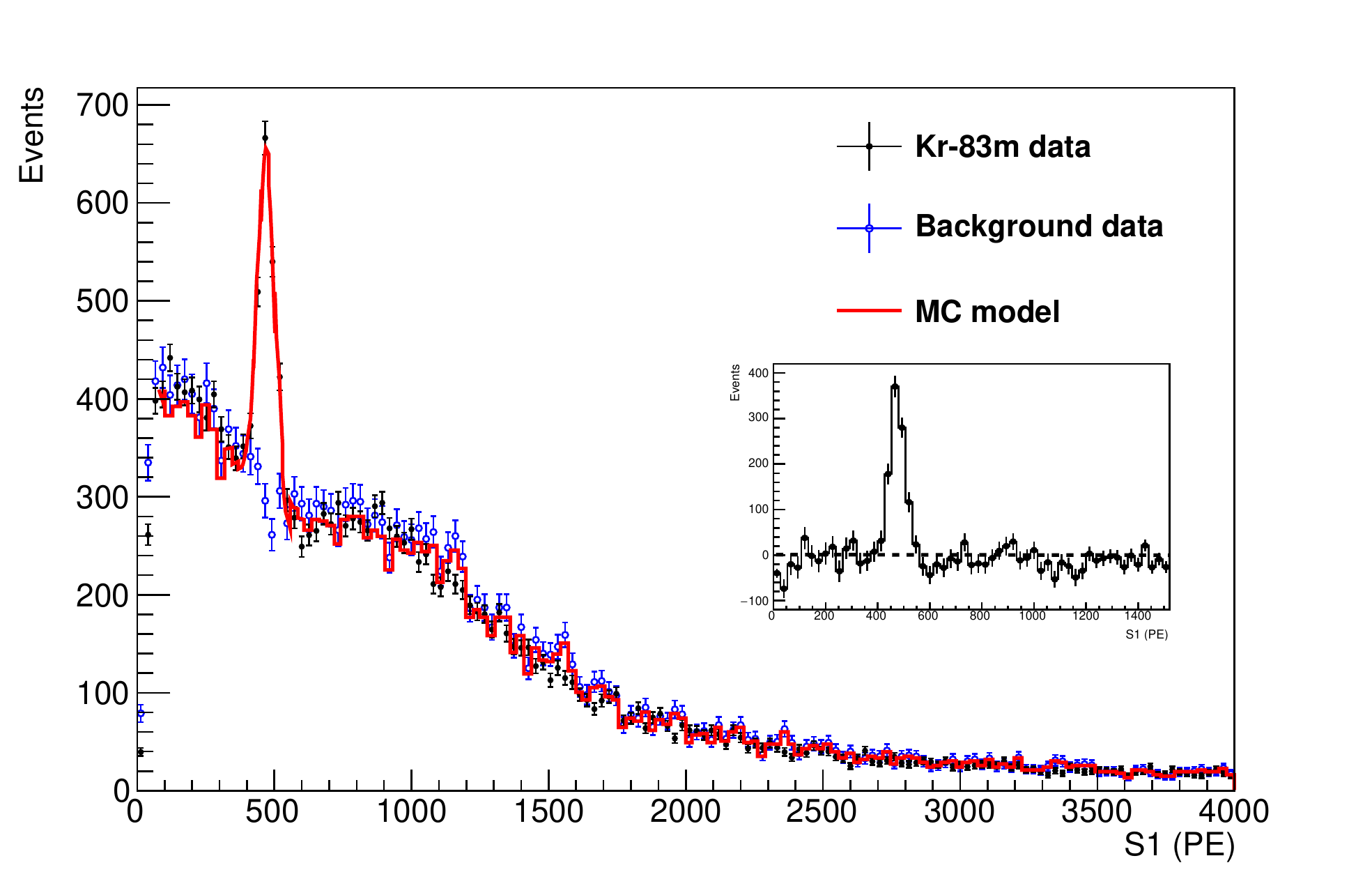}
\caption[\Kr\ distribution]{Measured S1 distribution from a \Kr\ calibration run with $\edrift = 183$\,V/cm, superimposed with the normalised environmental background and the best fit to a template (Monte Carlo signal + background data) that accounts for detector resolution. The inset shows the background-subtracted spectrum.}
\label{fig:s1-kr}
\end{figure}

\begin{figure*}[tp]
\centering
\includegraphics[trim=0.3cm 0 1.2cm 0.8cm,clip,height=0.66\columnwidth]{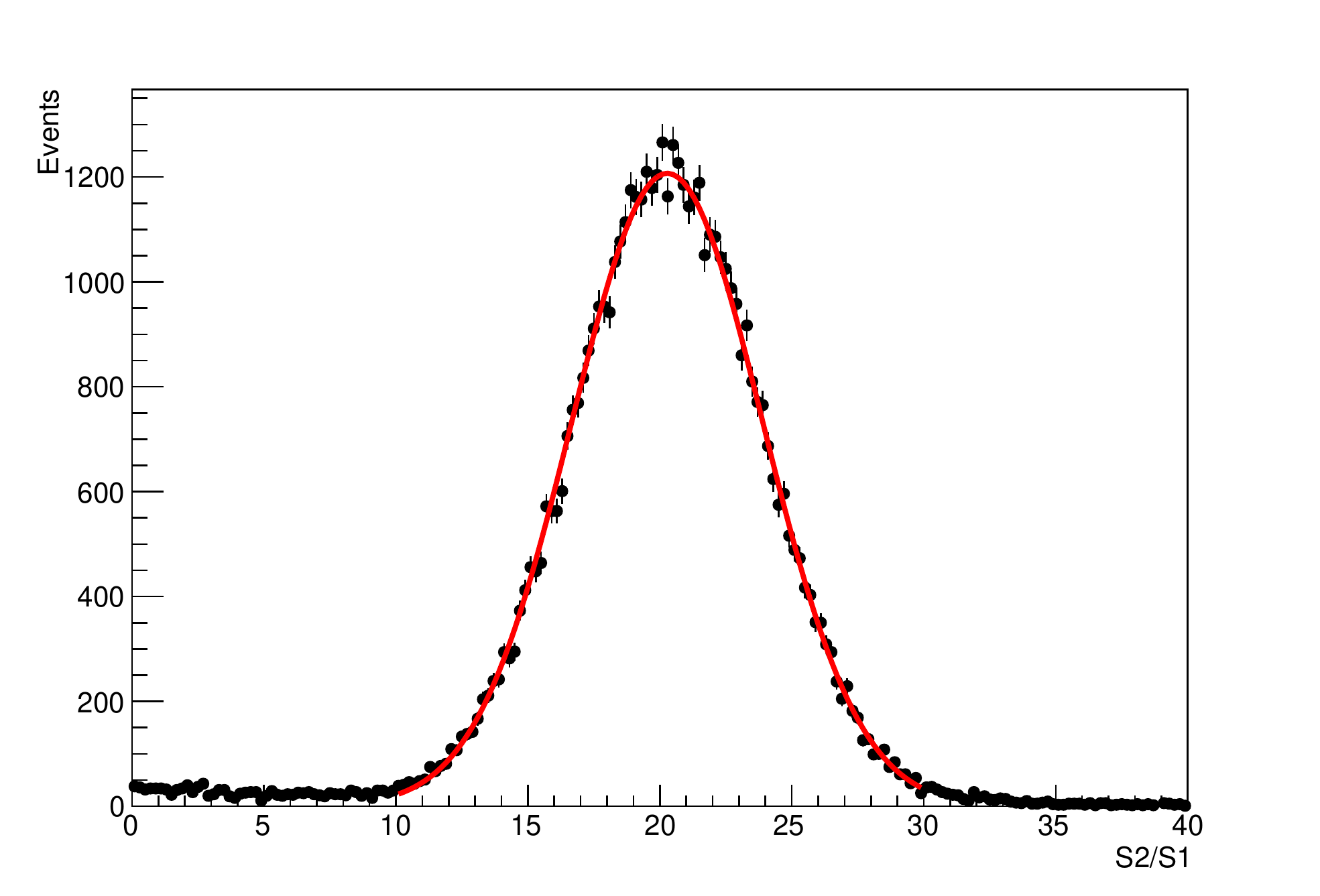}
\hspace{0.3cm}
\includegraphics[trim=0.3cm 0 1.2cm 0.8cm,clip,height=0.66\columnwidth]{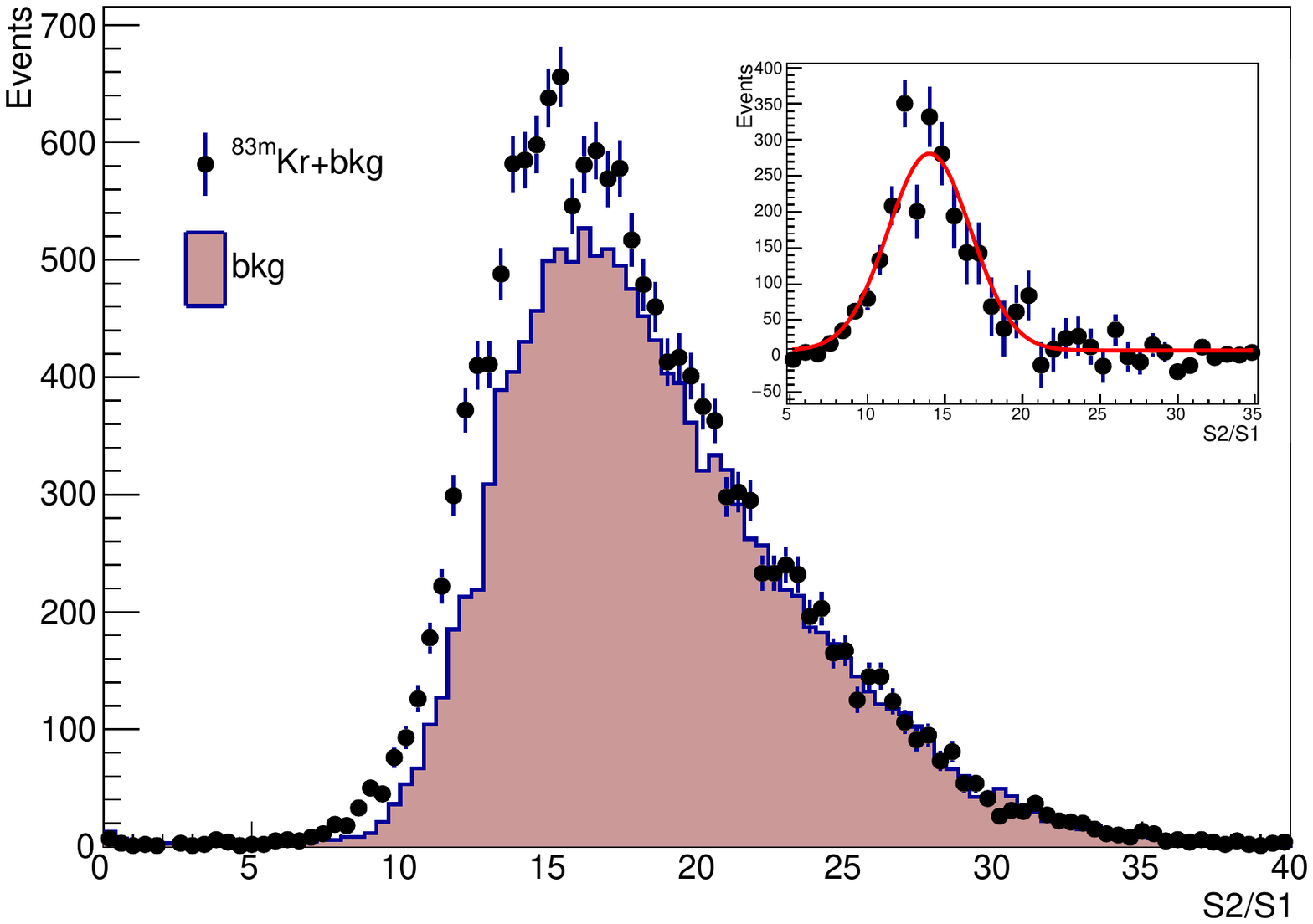}
\caption[S2/S1 distribution from \Am\ and \Kr\ calibrations in double phase at $\edrift = 183$\,V/cm]{Distribution of S2/S1 for \Am\ (\textit{left}) and \Kr\ (\textit{right}) measured at $\edrift = 183$\,V/cm.  
  The events are selected via S1 to only include the full-energy peak.
  For \Am, a Gaussian fit is superimposed to the data histogram to obtain the mean and width (rms) of the distribution. 
 For \Kr, the environmental background, normalised on the sideband (25--40\,PE) of the \Kr\ peak, is shown as a solid histogram. The background-subtracted \Kr\ signal distribution is shown in the inset, together with the asymmetric Gaussian function that best fits the data.}
\label{fig:s2overs1}
\end{figure*}

%ER
%
The internal \Kr\ source is a short-lived ($T_{1/2}= 1.83$\,h) progeny of $^{83}$Rb ($T_{1/2}= 86.2$\,days); being
a radioactive gas, \Kr\ can be diffused uniformly within the \LAr\ and \TPC\ using the procedure developed 
%for \DSf\,
in\,\cite{Agnes:2015gu}. \Kr\ decays by Isomeric Transition (IT), emitting two
monoenergetic conversion electrons (9 and 32\,keV) that are sometimes accompanied by associated fluorescence x-rays. Fig.\,\ref{fig:s1-kr} shows the S1 spectrum of \Kr\ measured at $\edrift = 183$\,V/cm. 
The S2/S1 distributions from  \Am\ and \Kr\ are shown in Fig.\,\ref{fig:s2overs1}, superimposed with a Gaussian fit. The S2/S1 resolution, calculated as the ratio of $\sigma/\mu$ from the fit, is $(17.9 \pm 0.1)$\% for \Am. In the case of \Kr, the data indicate a larger width ($\sigma/\mu \sim 25$\%), which is likely due to the signal being composed of many electrons and x-rays, producing larger fluctuations in the ionisation signal.
%Fig.\,\ref{fig:S2S1correlation} shows the S1-S2 correlation from \Am.  \color{black}
%
%
%\begin{figure}
%\centering
%\includegraphics[trim=0.15cm 0 0cm 0.6cm,clip,width=0.95\columnwidth]{figure/S2vsS1_scatterplot.pdf}
%\caption[S2 vs.\ S1 from a \Am\ calibration in double phase at $\edrift = 183$\,V/cm ]{Two-dimensional 
%S2 vs.\ S1 scatter plot for \Am\ measured at $\edrift = 183$\,V/cm. The population at $S1 \sim 500$\,PE 
%and $S2 \sim 11 \cdot 10^3$\,PE corresponds to the full-energy deposition of the prominent 60-keV $\gamma$-ray from \Am.}
%\label{fig:S2S1correlation}
%\end{figure}
%

Nuclear recoils are very slow and have a much larger stopping power ($dE/dx$) than electron recoils, according to the Bethe formula. The higher ionisation density results in an enhanced probability of electron-ion recombination, and hence a larger S1 and smaller S2 signal. The difference in ionisation density also produces a different proportion of two excited states of the Ar dimer, that then emit scintillation light with widely different time constants (approximately 6\,ns and 1.6\,\textmu s), permitting powerful ER/NR discrimination based on the time profile of the S1 signal~\cite{Amaudruz:2016dq}. The fraction of scintillation light emitted by the shorter-lived excimer is higher for NRs relative to ERs and is therefore often used to discriminate between the two classes of events. The parameter $f_{prompt}$ is defined here as the fraction of scintillation light taking place in the initial 700\,ns. While the optimisation of this parameter is beyond the scope of this work, this simple definition results in a NR/ER separation better than $2 \sigma$ at the lowest energy considered (50\,PE), which is sufficient for the purposes of this study.

\begin{figure*}[tbph]
\centering
\includegraphics[trim=0.2cm 0 1.4cm 0.95cm,clip,height=0.78\columnwidth]{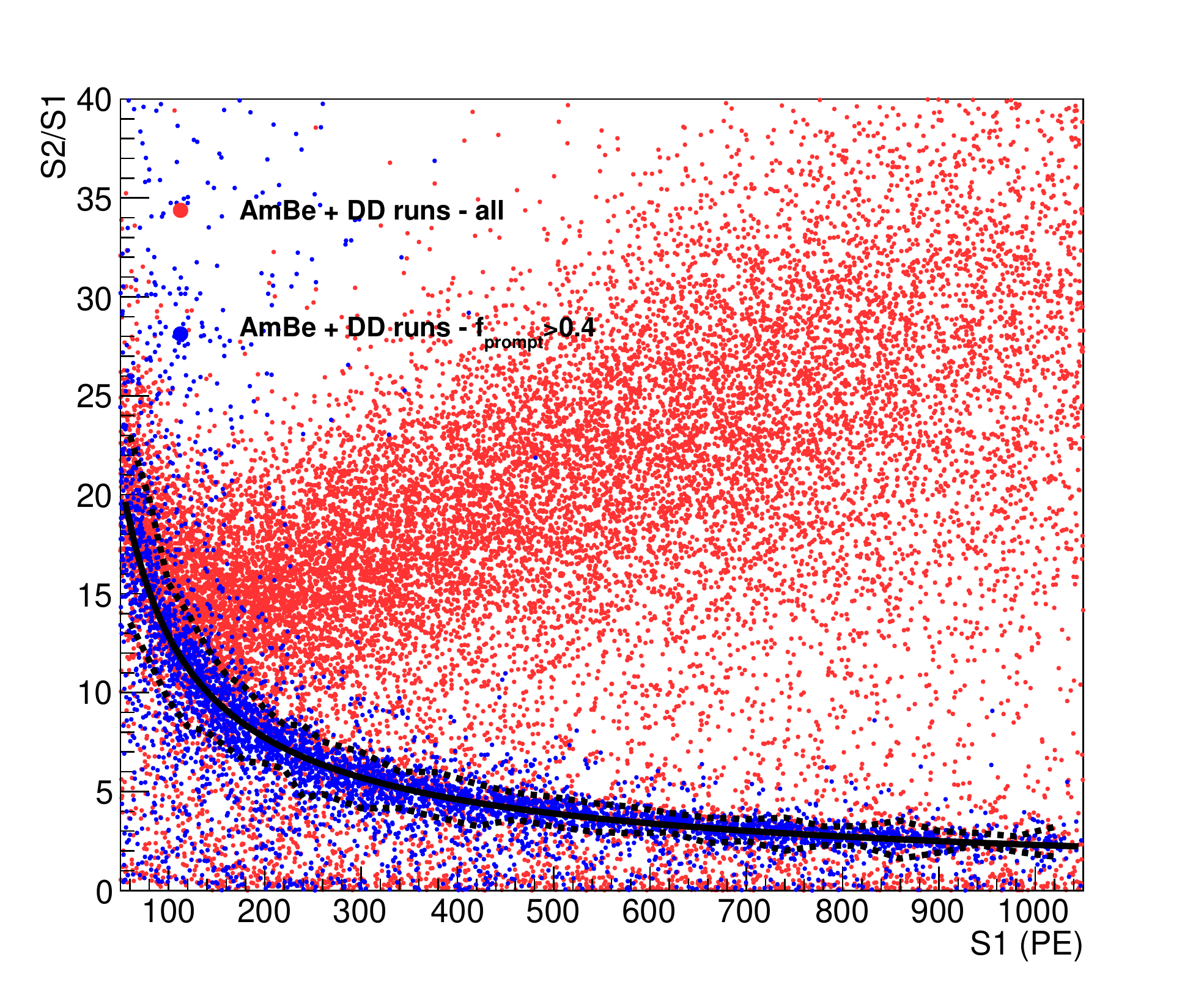}
\hspace{0.3cm}
\includegraphics[trim=0.3cm 0 1.4cm 0cm,clip,height=0.78\columnwidth]{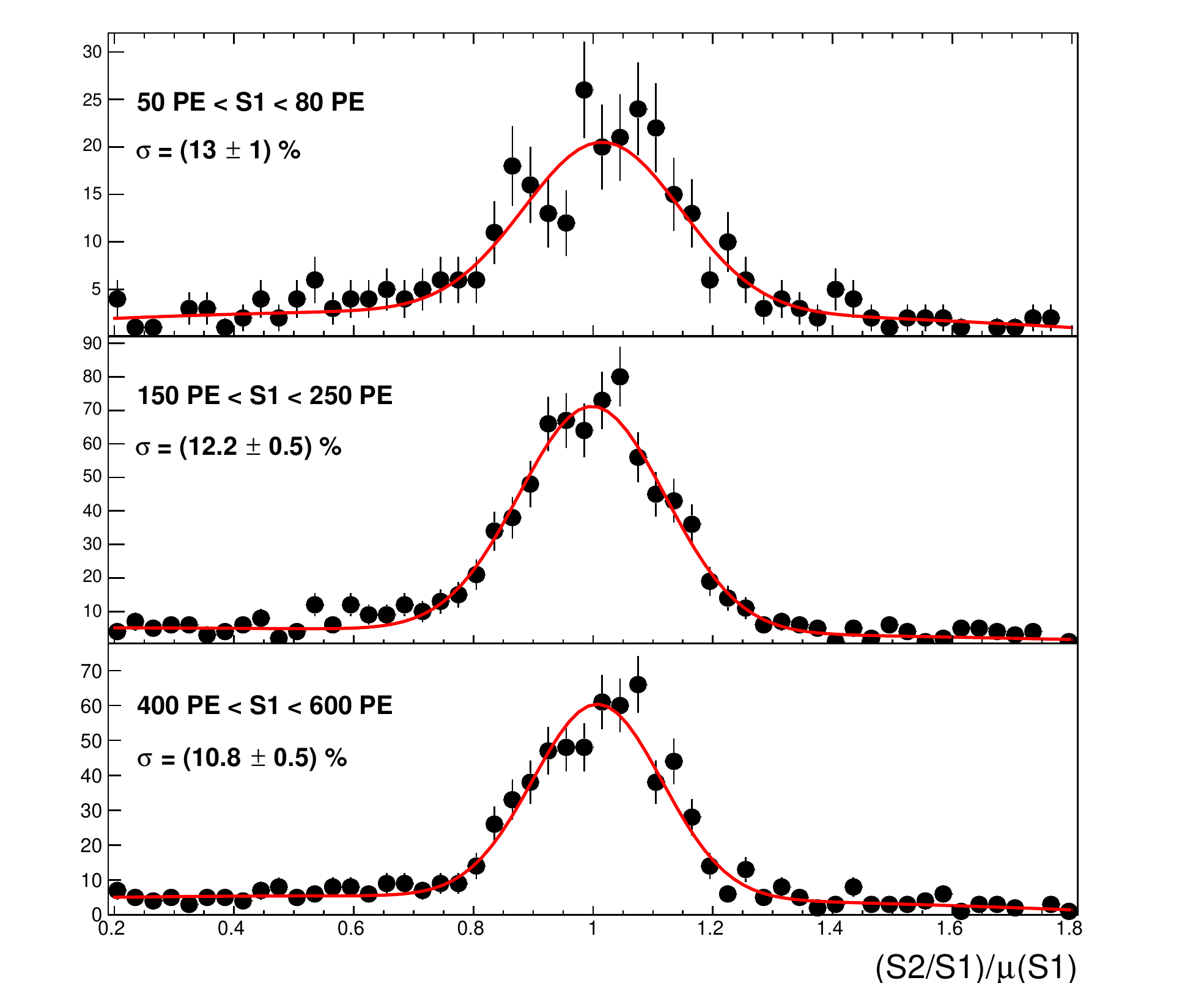}
\caption[S2/S1 distribution and mean value for NRs from DD+AmBEgun]{\textit{Left:} Distribution of NRs (blue points, $f_{prompt} > 0.4$) compared to all events (red points) in the S2/S1 vs.\ S1 plane for the AmBe + DD combined dataset. 
 The most probable values of S2/S1 as a function of energy, $\mu(S1)$, are represented by the black curve and the 90\% range is shown 
 by the black dashed curves. 
 \textit{Right:} Distribution of $\frac{(S2/S1)}{\mu}$ in three energy ranges: 50--80\,PE (top panel), 150--250\,PE (middle panel), and 400--600\,PE (bottom panel). } \label{fig:s2s1nr}
\end{figure*}

The \ReD\ \TPC\ was exposed to two neutron sources during the data campaign discussed here: an
AmBe source and a commercial deuterium-deuterium (DD) neutron generator~\cite{CHICHESTER2005753}. The AmBe source provides a broad spectrum of neutrons (up to $\sim$8\,MeV), resulting in $^{40}$Ar recoils up to 800\,keV. 
The DD neutron generator emits nearly monochromatic 2.5-MeV neutrons, which produce a NR spectrum up to 250\,keV; for radiation safety, its fluence was limited to $10^4$\,n/s over the entire solid angle. The two datasets provide consistent measurements and are combined hereafter in order to reduce statistical uncertainties. 
%The  scatter plot of the S2/S1 ratio vs. $f_{prompt}$ is shown in Fig.~\ref{fig:scatters2s1fprompt}, for 
%$S1 < 200$~PE (left) and $S1 > 200$~PE (right). 
%
%\begin{figure*}[tp]
%\centering
%\includegraphics[width=0.90\textwidth]{figure/Run1226_above_and_below200phe.pdf}
%\caption[S2/S1 vs. $f_{prompt}$]{
%  Distribution of S2/S1 vs. $f_{prompt}$ for a neutron-rich 
%  calibration taken with the DD-gun. Left: events with $S1<200$~PE, right: events with $S1>200$~PE.}
%\label{fig:scatters2s1fprompt}
%\end{figure*}
%
%It is clearly visible that different regions are populated, on the plane 
%S2/S1 vs. S1, according to $f_{prompt}$ values. 
The left panel of Fig.\,\ref{fig:s2s1nr} shows the S2/S1 ratio as a function of S1 for all single-scatter events with valid S1 and S2 signals, as well as for events compatible with a neutron-induced NR ($f_{prompt} > 0.4$). 
The NR band is clearly separated from the ER band above $\sim$200 PE. The mean value and width of S2/S1 are calculated in intervals of S1 with varying width between 20 and 40~PE using a model consisting of a Gaussian distribution summed with a linear function. The set of most probable values of the S2/S1 mean ($\mu$) is then fitted as a function of S1 using the empirical function $\mu(S1) = a(S1+b)^{c}$, shown as a black solid curve in the left panel of Fig.\,\ref{fig:s2s1nr}. The right panel of Fig.\,\ref{fig:s2s1nr} shows the distribution of (S2/S1)/$\mu(S1)$ for three different energy ranges: 50--80\,PE ($\sim$20--30\,keV$_{nr}$); 150--250\,PE, which includes the recoil energy used to benchmark the directional sensitivity of \ReD\ ($\sim$70 keV$_{nr}$); and 400--600\,PE, which includes the \Am\ peak shown in Fig.\,\ref{fig:s2overs1} for comparison. The dispersion of S2/S1 for NRs is calculated as the relative standard deviation ($\sigma/\mu$) of the Gaussian distribution and reported in Tab.\,\ref{tab:gf} for the three S1 ranges considered here, along with the corresponding tail fraction, defined as the fraction of events outside the interval $\mu \pm 1.96 \sigma$ (i.e. the 90\% quantile of a Gaussian distribution). In the highest energy range, the measured S2/S1 dispersion for NRs ($\sim$11\%) is considerably smaller than that observed for ERs ($\sim$18\%) in roughly the same S1 signal range. This difference is mostly due to a smaller amount of fluctuations in recombination for NRs than for ERs and, consequently, a better resolution in S2. The S2/S1 observable folds in the possible directional dependence of each of the samples, albeit integrated over a large angular range for the samples studied here. The measured S2/S1 dispersion in the energy range 150--200\,PE is 12\%, improving on previous results obtained by the SCENE collaboration~\cite{Cao:2014gns}, and sufficiently low to ensure that a potential directional effect with magnitude equal to that suggested by the results from SCENE would not be hidden by instrumental resolution.
In this regard, the performance of the \TPC\ reported here meets the requirements needed to accomplish the main goals of the \ReD\ experiment in the search for a directional effect due to columnar recombination in NRs.
%The fluctuations in the NR ionisation measurement are, as expected, significantly smaller than those observed for ERs and almost a factor of two smaller than those reported by the SCENE experiment. The performance of the \TPC\ is comparable, in this regard, to DarkSide-50 and appears to be sufficient for accomplishing the main goals of the \ReD\ experiment in the search for a directional effect due to columnar recombination in NRs.

%

\begin{table}[tbph]
\centering
\begin{tabular}{l | c  | c}
\hline
S1 range (PE) & $\sigma/\mu$ & tail fraction\\
\hline
%150 - 200 & 1.023(8) & 0.135(9)\\
%500 - 700 & 0.994(6) & 0.114(5)\\
50--80 & 0.13(1) & 35(3)\% \\
150--250 & 0.122(5) & 27(2)\% \\
400--600 & 0.108(5) & 33(2)\%\\\hline
\end{tabular}
\caption[fit parameters for S2/S1 resolution for NRs]{
Relative standard deviation ($\sigma/\mu$) and tail fraction from the fits of Fig.\,\ref{fig:s2s1nr}. The tail fraction is defined as the fraction of events outside the interval $\mu \pm 1.96 \sigma$, which
corresponds to the 90\% quantile of a Gaussian distribution. The significant fraction of events in the tails is due to a non-Gaussian S2 response, leakage from ER events, and pile-up. Only statistical uncertainties are reported here.}
\label {tab:gf}
\end{table}

\section{Dependence of scintillation and ionisation response on the drift field} \label{sec:s1s2}
In this section, measurements of scintillation and ionisation performed as
a function of the drift field ($\edrift$) are presented and discussed.

The passage of an ionising particle in a noble liquid produces both excitons ($N_{ex}$), which give rise 
to scintillation light, and electron-ion pairs ($N_i$)~\cite{Doke:2002oab,Doke:1988dp}. A fraction $R$ of 
these electron-ion pairs recombine, giving an additional contribution to the scintillation light, while the remaining unrecombined electrons can be drifted, multiplied and collected to form the ionisation signal. The total number of scintillation photons produced is
\begin{equation}
N_{ph} = \eta_{ex} N_{ex} + \eta_i R N_i,
\end{equation}
where $\eta_{ex}$ and $\eta_i$ are the efficiencies for excitons and recombined electron-ion pairs to 
produce a scintillation photon, respectively. In the absence of non-radiative quenching 
phenomena, both $\eta_{ex}$ and $\eta_i$ are expected to be equal to 
one\footnote{As in \cite{Cao:2014gns}, $\eta_{ex}$ and $\eta_i$ are defined here to not account for any internal or track-dependent quenching processes, e.g.\ Penning ionisation or biexcitonic Hitachi quenching~\cite{Hitachi:1992gq}, which instead affect $N_{ex}$ and $N_i$.}. Defining $\alpha$ as the ratio of excitons to electron-ion pairs ($\alpha \equiv N_{ex}/N_i$), the total S1 signal 
can be expressed as
\begin{equation}
\label{eqn:S1}
\textrm{S1} = g_1 N_{ph} = g_1 (\alpha + R) N_i, 
\end{equation}
where $g_1$ is the number of photoelectrons detected per scintillation photon emitted. The electrons surviving recombination contribute to the ionisation signal S2, which is expressed as
\begin{equation}
\textrm{S2} = g_2 (1-R) N_i, \label{eqn:S2}
\end{equation}
where $g_2$ is the S2 gain of the \TPC, i.e.\ the number of photoelectrons detected for each electron extracted from the liquid. The parameters $g_1$ and $g_2$ are intrinsic properties of the detector and account for light collection efficiency and photon detection efficiency of the SiPMs.

In the limit of full recombination ($R \to 1$), the average energy required for the production of one scintillation photon in \LAr\ can be written~\cite{Doke:2002oab} as
\begin{equation}
W_{ph}(\textrm{max}) = \frac{E}{N_{ex} + N_{i}} = \frac{W}{1+N_{ex}/N_{i}} = \frac{W}{1+\alpha},\label{eq:Wphmax}
\end{equation} 
where $W=E/N_i$ is the average energy required to produce an electron-ion pair. The values of $W$ and \Wphmax\ were measured to be $(23.6 \pm 0.3)$\,eV~\cite{Miyajima:1974zz} and $(19.5 \pm 1.0)$\,eV~\cite{Doke:2002oab}, respectively, using $\sim$1-MeV conversion electrons from a $^{207}$Bi source. These measurements correspond to a value of $\alpha=0.21\pm0.06$. %For NRs, the value of $\alpha$ is instead \color{red} dependent on recoil energy, ranging between 0.5 and 2 for recoils between 10 and 60\,keV$_{nr}$~\cite{Cao:2014gns,Alexander:2013ke,Cao:2015ks}. (which of these SCENE papers?) \color{black}
%$\sim$1~\cite{Doke:2002oab,Doke:1988dp}.

S1 and S2 signals are expected to be anti-correlated since they originate from competing scintillation and ionisation processes. Their relative balance depends on the recombination probability ($R$), which is affected by the presence of the drift field ($\edrift$) in the active region of the \TPC. In particular, for increasing $\edrift$, more electrons are swept away from the % high ion density 
interaction site and can therefore survive recombination.
%; this lower value of $R$ causes a reduction of the scintillation signal and an increase of the ionisation signal. 
The anti-correlation between S1 and S2 allows a determination of the gains $g_1$ and $g_2$, discussed in Sec.\,\ref{sec:doke}. The reduction of the scintillation light (``quenching'') measured in \ReD\ for increasing $\edrift$ and 
the fit of these data with an empirical model for the recombination probability are presented in Sec.\,\ref{sec:quenching}. 

\subsection{S1 and S2 correlation} \label{sec:doke}
The experimental gains $g_1$ and $g_2$ can be derived using the anti-correlation between S1 and S2 signals by following the procedure described in 
\cite{Cao:2014gns}. Defining the yields $Y_1=\mathrm{S1}/E$ and $Y_2=\mathrm{S2}/E$ as the number of photoelectrons measured in the S1 and S2 
signals, respectively, per unit of deposited energy, the anti-correlation between $Y_1$ and $Y_2$ can be written 
from Eqs.\,\ref{eqn:S1}, \ref{eqn:S2}, and \ref{eq:Wphmax} as
\begin{equation}
  Y_1 = \frac{g_1}{W_{ph}(\textrm{max})} - \frac{g_1}{g_2}Y_2.  \label{eq:dokeanticorr}
\end{equation}
%with \Wphmax\ as in Eq.\,\ref{eq:Wphmax}.
For this study the full absorption peak of the \Am\ source is taken from data collected in double-phase mode at $100 \leq \edrift \leq 1000$\,V/cm. Data collected at null field are 
not included here since any electrons surviving recombination, referred to as ``escape electrons''~\cite{Doke:2002oab}, are not drifted 
and remain undetected. To ensure a proper comparison between measurements taken at different values of $\edrift$, only events localised in the central part of the detector are considered, where the drift field is expected to be more uniform. The $x$-$y$ position of an event is approximately estimated as the centre of the SiPM in the top tile with the highest fractional S2 charge. Only events with a reconstructed $x$-$y$ position corresponding to one of the inner eight SiPMs of the top tile are selected. The correction due to leakage current in the SiPM (see Sec.\,\ref{sec:ser}) is also applied to the measured S1 and S2 yields. Systematic uncertainties on $Y_1$ and $Y_2$ are largely dominated by the leakage current correction, with sub-leading contributions from the MC templates and other fit-related uncertainties.

\begin{figure}[tbph]
\centering
\includegraphics[trim=0.2cm 0 1.4cm 0.6cm,clip,width=0.95\columnwidth]{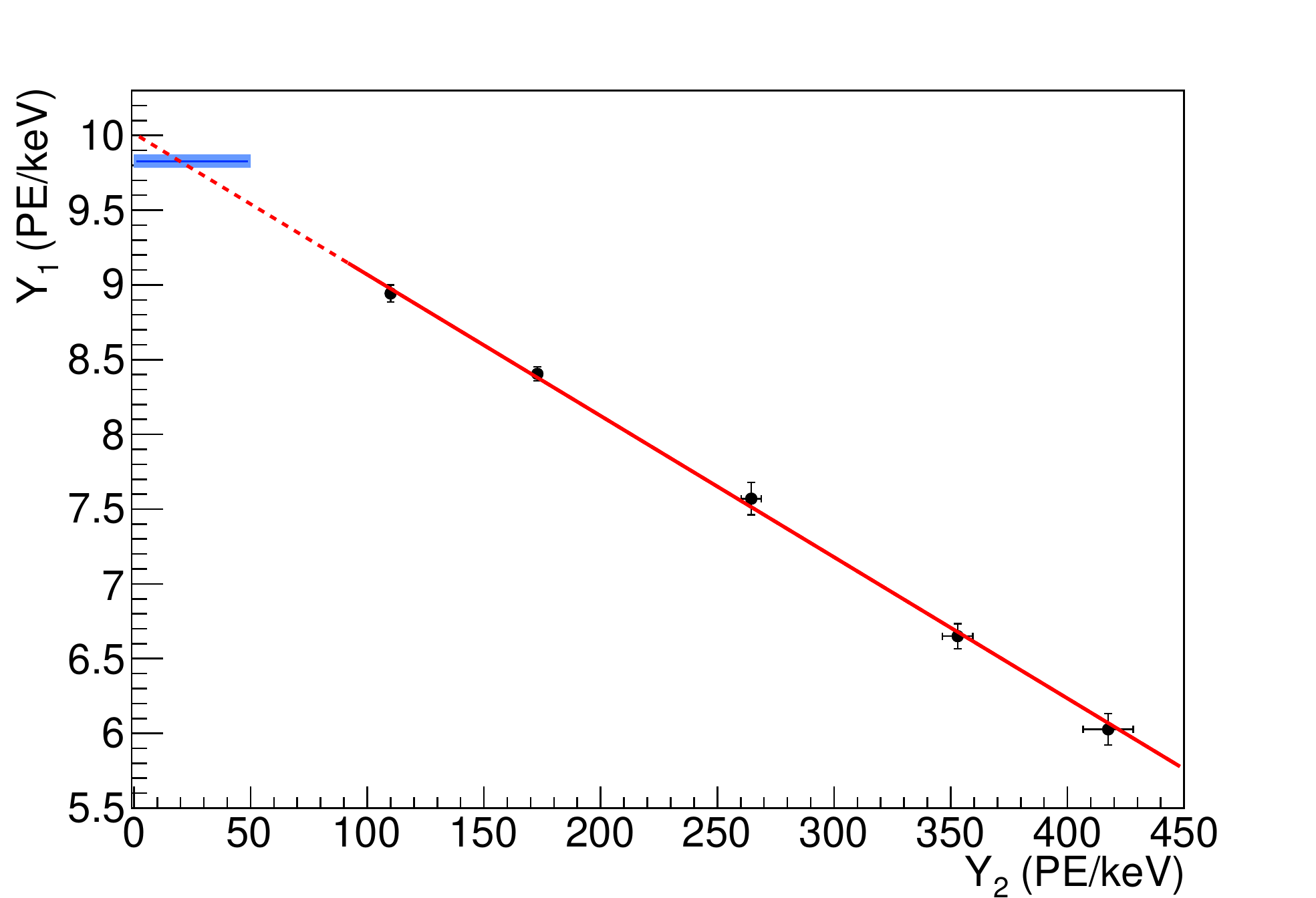}	
\caption[Doke-plot: S1-S2 anti-correlation for \Am\ measurements taken at different drift fields]{S1 vs.\  S2 yield for \Am\ calibration runs taken in double-phase mode at
%$\edrift = 92.1, 183.5, 393.8, 696.0$ and $995.7$\,V/cm.
$100 \leq \edrift \leq 1000$\,V/cm.
The red line is the best fit according to the model of Eq.\,\ref{eq:dokeanticorr}. %Assuming \Wphmax\ $= (19.5 \pm 1.0)$\,eV for 60-keV \Am\ $\gamma$-rays, the best fit values correspond to $g_1 = (0.1953 \pm 0.0015$)\,PE/photon and $g_2 = (20.67 \pm 0.66$)\,PE/e$^{-}$. 
The horizontal blue band shows the S1 yield measured at null field $Y_1^0 = (9.80 \pm 0.05_{\rm \,stat}$)\,PE/keV, where a measurement of $Y_2$ is not possible.
% due to the lack of a drift field. 
%due to the possible contribution of escape electrons.
}\label{fig:dokeplot}
\end{figure}

%Data with overlaid fit are shown in Fig.~\ref{fig:s1s2example} for the run at 183 V/cm.
% 
%\begin{figure}[tbph]
%\centering
%\includegraphics[width=0.45\textwidth]{figure/S1internal_fit_200.pdf}
%\includegraphics[width=0.45\textwidth]{figure/S2internal_fit_200.pdf}
%\caption[S1 S2 fits example]{S1 (left) and  S2 (right)  \Am\ peak at 183 V/cm with overlaid fit based on Monte Carlo template.} \label{fig:s1s2example}
%\end{figure}
The $Y_2$-$Y_1$ correlation obtained for $75 \leq \edrift \leq 1000$\,V/cm is shown in Fig.\,\ref{fig:dokeplot}, along with the best fit to the linear model of Eq.\,\ref{eq:dokeanticorr}. Assuming \Wphmax\ $= (19.5 \pm 1.0)$\,eV for 59.54-keV \Am\ $\gamma$-rays, the best fit parameters are $g_1 = (0.1953 \pm 0.0015)$\,PE/photon and $g_2 = (20.67 \pm 0.66)$\,PE/e$^{-}$.  The horizontal blue band in Fig\,\ref{fig:dokeplot} shows the light yield measured at null field, 
$Y_1^0 = (9.80 \pm 0.05_{\rm \,stat}$)\,PE/keV; $Y_2^0$ is not measured in this case due to the absence of a drift field. The predicted S1 yield at $Y_2=0$ from extrapolating the linear fit, i.e.\ under the implicit assumption that all electrons recombine at null field, is ($10.01 \pm 0.08_{\rm \,stat}$)\,PE/keV. The ratio $\eta$ between $Y_1^{0}$ and the value extrapolated to $Y_2=0$ is related to the fraction of electrons escaping recombination at null field by $\chi = (1+\alpha)(1-\eta)$~\cite{Doke:2002oab,Hitachi:2019nce}. The value of $\eta \sim 0.98$ calculated here for 59.54-keV $\gamma$-rays indicates that the contribution from escape electrons is limited to a few percent. 

The measured gains of the \ReD\ \TPC\ are $g_1 = (0.195 \pm 0.018_{\rm \,stat+sys})$\,PE/photon and
$g_2 = (20.7 \pm 1.6_{\rm \,stat+sys})$\,PE/e.
The systematic uncertainties on $g_1$ and $g_2$ are dominated by the uncertainty on 
\Wphmax, which contributes 5.1\%.
The measured value of $g_2$ is in good agreement with the independent estimate based on ``echo'' events discussed in~\ref{sec:echoes}. The measured value of $g_1$ in \ReD\ can be compared with that from DarkSide-50, $0.157 \pm 0.001$\,PE/photon~\cite{Agnes:2017grb}, and that from SCENE, $0.104 \pm 0.006$ PE/photon~\cite{Cao:2014gns}. The higher
value of $g_1$ achieved in \ReD\ is driven mostly by the better optical coverage of the \ReD\ \TPC\ and higher detection efficiency of the SiPMs with respect to photomultipliers. The S2 gain of  \ReD\ is comparable to that of DarkSide-50, $23 \pm 1$\,PE/e$^{-}$~\cite{Agnes:2018ves}, and significantly higher than that of SCENE, $3.1 \pm 0.3$\,PE/e$^-$~\cite{Cao:2014gns}. Based on the measured values of $g_1$ and $g_2$, the performance of the \TPC\ reported here satisfies the requirements needed to achieve the scientific goals of the \ReD\ experiment.

\subsection{Scintillation quenching and charge yield vs.\ $\edrift$} \label{sec:quenching}
The S1 and S2 response of the \ReD\ \TPC\ was studied as a function of the drift field using 
a dedicated set of \Am\ measurements taken at $0 \leq \edrift \leq 1000$\,V/cm in single- and double-phase
mode, together with $^{133}$Ba and $^{137}$Cs measurements taken in double-phase mode at 183 and 693\,V/cm.
From Eq.\,\ref{eqn:S1}, the ratio S1/S1$_0$ between the scintillation yield at a given value of $\edrift$ and that at null field can be expressed as%~\cite{Doke:2002oab}
\begin{equation}
\textrm{S1}/\textrm{S1}_0 = \frac{\alpha + R(dE/dx, \edrift)}{\alpha + R_0(dE/dx)} = \frac{\alpha + R(dE/dx, \edrift)}{\alpha + 1 -\chi(dE/dx)} \label{s1_over_s10},
\end{equation}
where the recombination probability ($R$) depends on the stopping power ($dE/dx$) and on the drift field ($\edrift$), $\chi$ is the fraction of escaping electrons at null field, and $R_0 = 1 - \chi$ is the recombination probability at null field.

In this work, $R$ is parameterised as a function of $dE/dx$ and $\edrift$ 
according to the Doke-Birks empirical recombination model~\cite{Doke:2002oab,Doke:1988dp}, modified to account for the observed dependence on $\edrift$~\cite{Agnes:2018mvl} and further extended here: 
\begin{equation}
R = \frac{A e^{-D_1 \edrift} \ dE/dx}{1+ B \ dE/dx} + C e^{-D_2 \edrift}, \label{eq:R-definition}
\end{equation}
where the constant $B$ is defined such that $R \to 1$ for highly-ionising particles ($dE/dx \to \infty$) at any value of $\edrift$:
\begin{equation}
B = \frac{Ae^{-D_1 \edrift}}{R(dE/dx \to \infty)-Ce^{-D_2 \edrift}}.
\end{equation}  
The parameterisation of Eq.\,\ref{eq:R-definition} is valid for tracks approximating a long column of electron-ion pairs (above 
$\sim$50\,keV). The second term represents `geminate', or Onsager, recombination~\cite{Doke:1988dp,Thomas:1987ek}, 
which occurs when an ionisation electron recombines with its parent ion, while the first 
term represents `volume' recombination, which occurs when a wandering ionisation electron is 
captured by an ion other than its parent. Following~\cite{Doke:2002oab}, the recombination at null field ($R_0$) can be expressed as
\begin{equation}
\label{eq:R0-definition}
R_0 = 1 - \chi = \eta (1+\alpha) -\alpha,
\end{equation}
where $\eta$ % = (1 + \alpha - \chi)/(1 + \alpha)$ 
is as defined in Sec.\,\ref{sec:doke}. Here it is parameterised as a function of $dE/dx$~\cite{Doke:2002oab}: %accounting for non-unity recombination probability
\begin{equation}
\eta(dE/dx) = \frac{A_0  \ dE/dx}{1+ B_0 \ dE/dx} + C_0, \label{eq:eta-definition}
\end{equation}
with the condition $B_0 = A_0/(1-C_0)$ imposed in order to guarantee that both $R_0$ and $\eta$ approach unity as $dE/dx \to \infty$.

A combined $\chi^2$ fit to several input datasets is performed, including measurements of S1/S1$_0$ taken by \ReD\ in single- or double-phase mode using \Am, $^{133}$Ba or $^{137}$Cs sources at $50 \leq \edrift \leq 1000$\,V/cm, as well as measurements of the S2 yield with an \Am\ source at $75 \leq \edrift \leq 1000$\,V/cm. Measurements of S1/S1$_0$ by the ARIS collaboration~\cite{Agnes:2018mvl} using $\gamma$ sources and single Compton scatters are also included in the analysis. The fit implements the model described in Eqs.\,\ref{eqn:S2}, \ref{s1_over_s10}, \ref{eq:R-definition}, \ref{eq:R0-definition} and \ref{eq:eta-definition} keeping a total of nine free parameters: $A$, $C$, $D_1$ and $D_2$ from the recombination probability parameterisation of Eq.\,\ref{eq:R-definition}; $A_0$ and $C_0$ from the parameterisation for the scintillation efficiency at null field of Eq.\,\ref{eq:eta-definition}; the excitation to ionisation ratio ($\alpha$); the ionisation work function in \LAr\ ($W$); and the S2 gain ($g_2$). The latter two parameters, $W$ and $g_2$, are additionally constrained via Gaussian penalty terms to the values $(23.6 \pm 0.3)$\,eV and $(21.0 \pm 1.3)$\,PE/e$^{-}$, respectively, taken from \cite{Miyajima:1974zz} and from the independent measurement described in\,\ref{sec:S3appendix}. %:$g_2 = (21.0 \pm 1.3)$\,PE/e$^{-}$. 
%Similarly, a Gaussian penalty term is added to constrain the ionisation work function ($W$) using the value $(23.6 \pm 0.3)$\,eV from~\cite{Miyajima:1974zz}. 
Finally, the $Y_2$-$Y_1$ correlation of Eq.\,\ref{eq:dokeanticorr} is recast to depend on the ratio 
$g_2/W_{ph}(\textrm{max})$ and subsequently on the combination $g_2(1+\alpha)/W$ using Eq.\,\ref{eq:Wphmax}. The fit of Sec.\,\ref{sec:doke} is then used to provide an additional constraint on $g_2(1+\alpha)/W = (1040\pm40)$\,PE/(e$^{-} \cdot \,$eV). The electron
$dE/dx$ used in Eq.\,\ref{eq:R-definition} is taken from the ESTAR database\,\cite{estar}, based on\,\cite{ICRU37}.

\begin{figure}
\centering
\includegraphics[trim=0.3cm 0 1.25cm 0.5cm,clip,width=0.95\columnwidth]{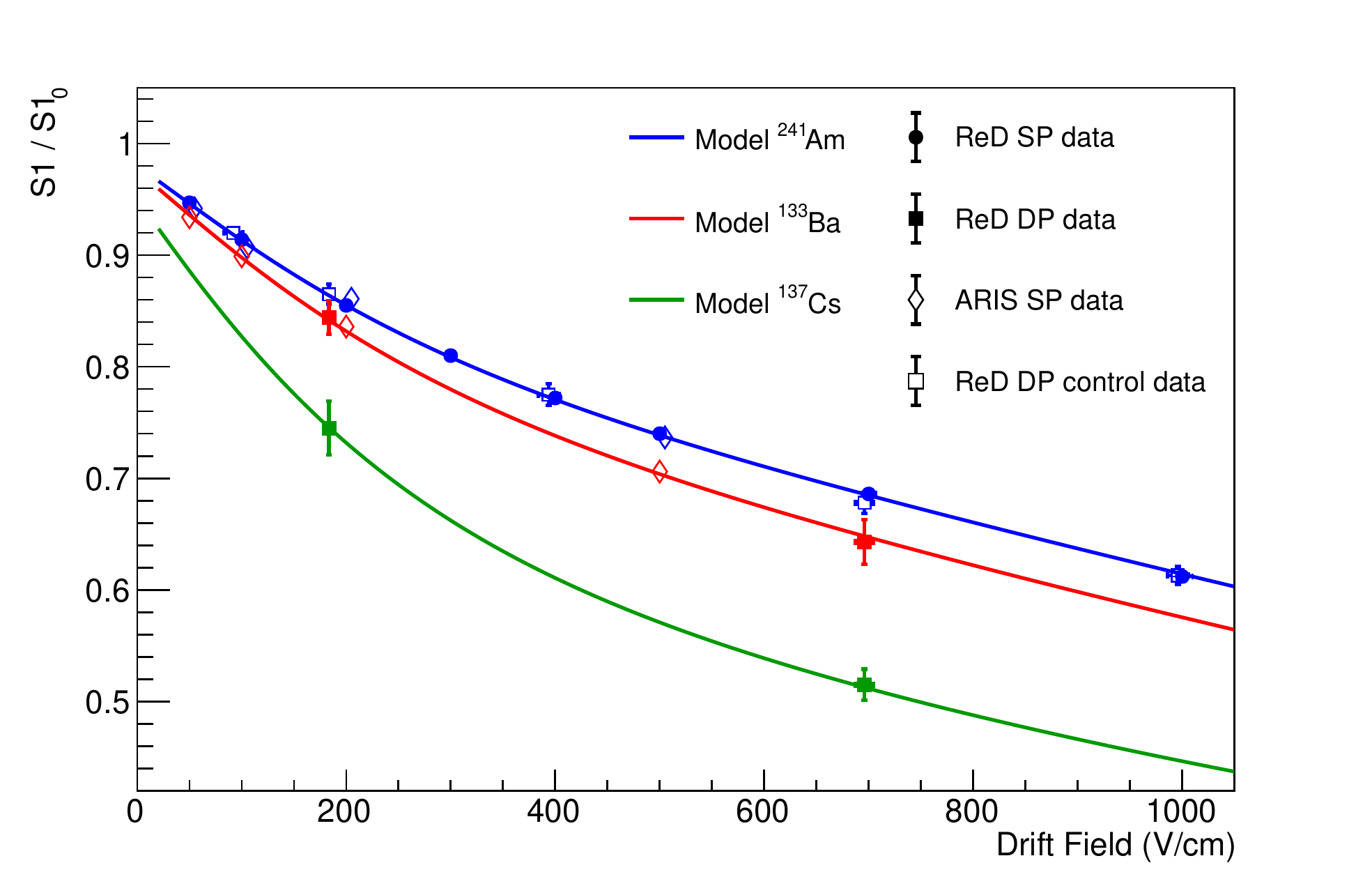} 
\caption[S1 quenching vs. drift field]{Ratio of S1 signal at $\edrift > 0$ to that at null field ($S1/S1_0$) as a function of $\edrift$, from \ReD\ single- and double-phase (DP) measurements with \Am, $^{133}$Ba and $^{137}$Cs sources, as well as ARIS single-phase (SP) measurements. Data points from ARIS are only available up to 500\,V/cm. The curves show the results of a combined fit using both S1 and S2 data to the modified Doke-Birks parameterisation of Eqs.\,\ref{s1_over_s10} and \ref{eq:R-definition} for an energy deposit of 59.54\,keV (\Am, blue), 81\,keV ($^{133}$Ba, red), and an average energy deposit of 400\,keV ($^{137}$Cs Compton peak, green). Fit parameters are given in Tab.\,\ref{tab:DB} (`\ReD\ + ARIS'). \ReD\ \Am\ double-phase data (blue empty squares) are kept as a control sample and not used in the fit.}
\label{fig:s1s0}
\end{figure}

\begin{figure}
\centering
\includegraphics[trim=0 0 0.3cm 0.5cm,clip,width=0.98\columnwidth]{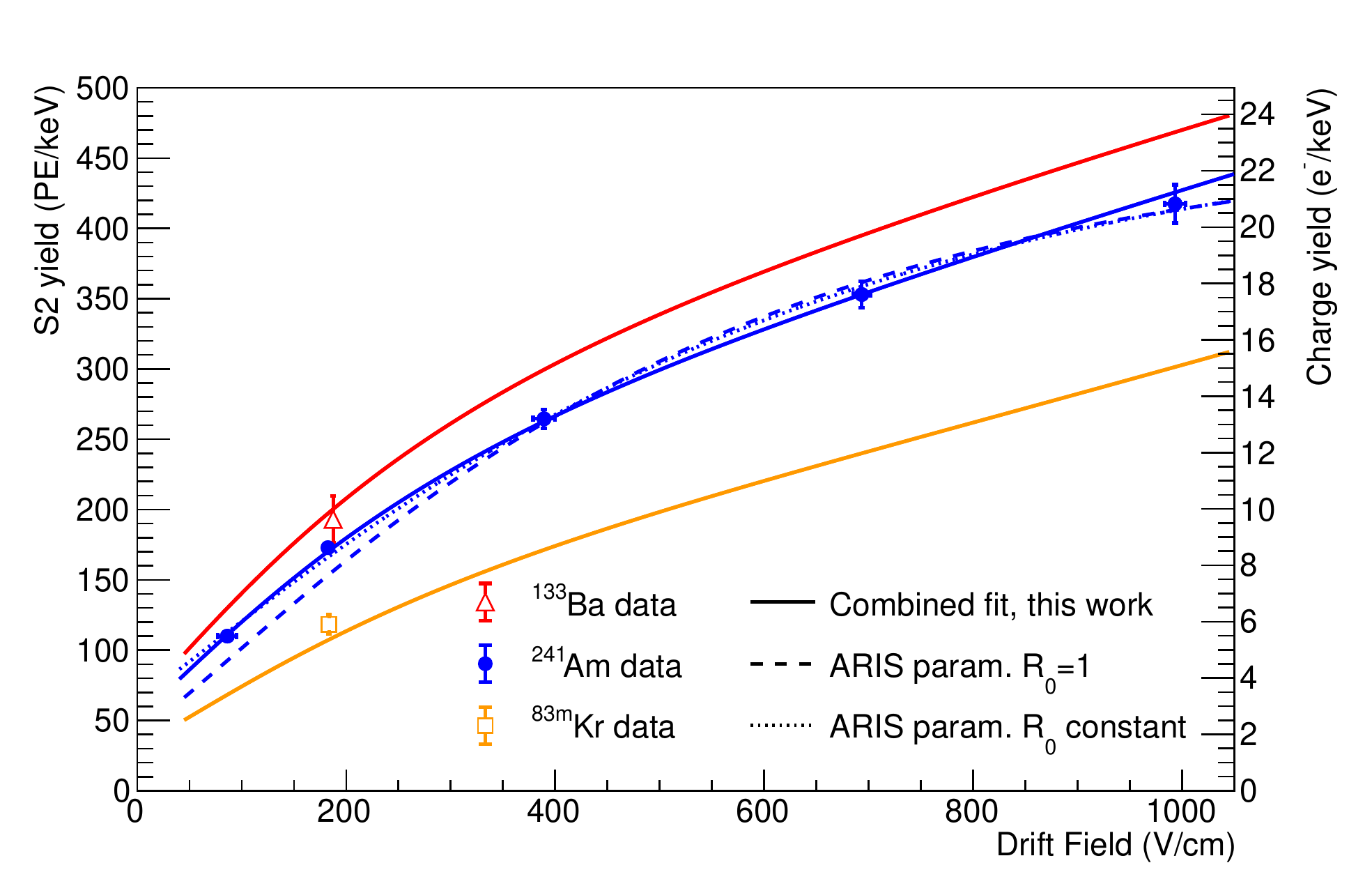} % V2
\caption[\Qy]{S2 yield (left axis) and charge yield (right axis) vs.\ $\edrift$, from \ReD\ double-phase measurements with \Am, \Kr\ and $^{133}$Ba sources. The solid lines correspond to the combined `ARIS + ReD' fit described in the text, with fit parameters given in Tab.\,\ref{tab:DB}. The dashed line shows the best fit to the recombination parameterisation employed by the ARIS collaboration~\cite{Agnes:2018mvl} ($D_1 = 0$, $R_0 =1$), while the dotted line lifts the constraint of full recombination at null field. Data taken with \Kr\ and $^{133}$Ba sources are kept as a control sample and not used in the fit.}\label{fig:qyvsfield}
\end{figure}

Results from this combined fit are reported in Tab.\,\ref{tab:DB}, while data for S1/S1$_0$ and the S2 yield are shown in Figs.\,\ref{fig:s1s0} and \ref{fig:qyvsfield}, respectively, with the fit overlaid. The vertical axis on the right-hand side of Fig.\,\ref{fig:qyvsfield} shows the charge yield (\Qy), a detector-independent quantity defined as the average number of electrons released per unit of deposited energy and calculated as $Y_2/g_2$, using the fitted value for $g_2$. The fit has 50 degrees of freedom and a ($\chi^2$-based) p-value of 74\%, once published uncertainties on the ARIS single-Compton dataset are inflated by 50\%. If the original uncertainties are used instead, the fitted parameters remain stable within one standard deviation, but the p-value drops to 0.3\%.

The set of \ReD\ S1/S1$_0$ measurements taken in double-phase mode with \Am\ is kept as a control sample and not used in the combined fit. The control data, shown in Fig.\,\ref{fig:s1s0} as blue empty squares, are in excellent agreement with the model prediction. Similarly, \ReD\ data taken using \Kr\ and $^{133}$Ba sources %at $\edrift=183$\,V/cm 
are not used in the combined fit and instead used to cross-check model predictions of the charge yield, shown in Fig.\,\ref{fig:qyvsfield} as ochre and red solid lines, respectively. The two data points are in reasonable agreement with the prediction, although the \Kr\ signal is composed of low energy electrons (9 and 32 keV) and therefore outside the range of validity of Eq.\,\ref{eq:R-definition}.
%\textcolor{red}{Higher energy source data are not used for 
%ionization measurement due to  saturation in the SiPM front-end electronics caused by the relatively 
%larger signals.}.

%The estimate of the S2 gain ($g_2$) from the combined fit is compatible with the independent measurement from ``echo'' events, described in App.\,\ref{sec:S3appendix}. 
The S1 gain ($g_1$) can also be derived by rescaling the fit of Sec.\,\ref{sec:doke} with the newly fitted value for $W_{ph}(\textrm{max})=W/(1+\alpha)$. These measurements have a smaller uncertainty with respect to those previously reported in Sec.\,\ref{sec:doke} and represent the final assessment of the \ReD\ TPC S1 and S2 gains:
\begin{eqnarray}
g_1 & = & (0.194 \pm 0.013_{\rm \,stat+sys})\,{\rm PE/photon} \nonumber \\
g_2 & = & (20.0 \pm 0.9_{\rm \,stat+sys})\,{\rm PE/e}^-. \label{eq:g1-g2-fit}
\end{eqnarray}
% constant-eta (shall we drop paragraph & line?)
A measurement of the escape probability at null field for electron recoils induced by 59.54-keV \Am\ $\gamma$-rays is derived from the combined fit: $\chi(^{241}\mathrm{Am})= 1 - R_0 = (3.6\pm0.6)\%$. Since the variation in $\eta$ over the range of energies discussed here is expected to be below 3\%, a fit with a constant escape probability as a function of energy ($A_0 \equiv 0$) is also performed. This fit is consistent with the nominal fit for all parameters of interest and has a comparable p-value.

An alternative fit is performed using the recombination parameterisation employed by the ARIS collaboration~\cite{Agnes:2018mvl}, which assumes that the volume recombination term is field-independent ($D_1 = 0$) and that all electrons eventually recombine within the experimental observation time ($R_0=1$). The resulting fit, shown in Fig.\,\ref{fig:qyvsfield} as a dashed line, is in disagreement with the data points at low $\edrift$ and returns a p-value below 0.1\%.
If the assumption $R_0=1$ is relaxed, the ARIS parameterisation gives a good description of the \Am\ S2 data (see Fig.\,\ref{fig:qyvsfield}, dotted line), but fails to reproduce the behaviour of S1/S1$_0$ for the higher energy source data. The parameterisation of Eq.\,\ref{eq:R-definition} is therefore better suited for describing new measurements from \ReD\ at $\edrift > 500$\,V/cm, as well as the combined analysis of scintillation and ionisation signals presented here.

Assuming 1) fully efficient electron extraction to the gas phase and 2) the absence of non-radiative quenching mechanisms, the combined S1 and S2 analysis can be used to constrain the total number of quanta (exciton and ion-electron pairs) and the excitation to ionisation ratio ($\alpha$) for low energy ER events, given a value of the ionisation work function ($W$). Using the value of $W$ from \cite{Miyajima:1974zz}, the fit returns $\alpha=0.25\pm0.05$, compatible with values in the literature obtained for various energy regimes with different measurement techniques~\cite{Doke:2002oab,Doke:1988dp}. This value of $\alpha$ corresponds to a value of $W_{ph}(\textrm{max})= W/(1+\alpha) = (18.9 \pm 0.8)$\,eV. 
%, with the assumption that the 
%work function $W$ measured for 1~MeV electrons also applies at lower energies.
%In order to test the validity of the first or second assumption, the combined fit is performed without the additional constraint on $g_2$ from the measurement of ``echo'' events or without the constraint on the ionisation work function ($W$), respectively. The results are reported in Tab.\,\ref{tab:DB} and are consistent with the nominal fit within statistical uncertainties. In the latter case, the resulting fit gives a good description of the data, but returns a slightly larger value of $W$, $(26.6 \pm 0.2)$\,eV, which is compensated by a higher escape probability (see Tab.\,\ref{tab:DB}). 
% MR changed after Walter comments
In order to test the validity of the first assumption above, the combined fit is performed without the additional constraint on $g_2$ from the measurement of ``echo'' events, as in this case $g_2$ would not include possible inefficiencies in the extraction of electrons from the liquid. Removing the constraint on $W$ allows testing the impact of the second assumption, since the total number of quanta and the number of ion-electron pairs become independent variables. The results are reported in Tab.\,\ref{tab:DB} and are consistent with the nominal fit within statistical uncertainties. In the latter case, the fit gives a good description of the data, but returns a slightly larger value of $W$, $(26.6 \pm 0.2)$\,eV, which is compensated by a higher escape probability (see Tab.\,\ref{tab:DB}).

%In this case the sensitivity to the overall number of quanta generated in the ER events is retained using the known ionization amplification factor to recover the true number of electrons involved. 
%The model is able to reproduce the data perfectly, but returns a slightly larger $W$ value of $26.6 \pm 0.2\,{\mathrm eV}$, which is compensated by a higher escape probability. 
%Therefore this analysis is not able to discriminate against this latter scenario.
%
\begin{table*}[tb]\small
\centering
\begin{tabular}{l | cccccccc}
\hline
\hline
Data & A & C & D$_1$ & D$_2$ & A$_0$ & C$_0$ & $\alpha$  & $g_2$\\
 & cm/MeV& & cm/kV & cm/kV &cm/MeV & & & PE/$e^{\tiny -}$\\
\hline
%ARIS~\cite{Agnes:2018mvl} & 0.25(2) & 0.77(1)& -- & 0.0035(3) & 0 & 1 & 0.21 & --\\
\ReD\ + ARIS                              & 0.49(5) & 0.56(2)& 1.1(1) & 5.3(7) & 0.1(1) & 0.87(9) & 0.25(5)  & 20.0(0.9)\\ 
\hline
\ReD\ only                                   & 0.50(6) & 0.56(2)& 1.1(1) & 5.8(9) & 0.7(1.2) & 0.64(30) & 0.26(5) & 19.8(0.9)\\
%\ReD\ + ARIS  $R_0$ constant   & 0.52(5) & 0.65(2)& 1.1(1) & 0.0057(7) & --  & 0.97(1) & 0.29(5) & 18.4(0.7)\\ 
\ReD+ARIS, no $g_2$ constraint      & 0.48(6) & 0.56(2)& 1.1(1) & 5.4(7) & 0.1(2)  & 0.87(13) & 0.27(8) & 19.6(1.4)\\ 
\ReD+ARIS, no $W$ constraint  & 0.45(6) & 0.56(2)& 1.1(1) & 5.3(7) &0.02(3) & 0.92(5) & 0.31(8) & 21.0(1.3)\\ 
\hline
\hline
\end{tabular}
\caption[Fit parameters for the Doke-Birks recombination model]{Fit parameters for the modified Doke-Birks 
recombination model of Eq.\,\ref{eq:R-definition} based on \ReD\ and ARIS data. The value of the ionisation work function ($W$) is constrained to $(23.6 \pm 0.3)$\,eV~\cite{Miyajima:1974zz} for all fits, except for that labeled `no $W$ constraint', for which the fit returns a value $(26.6 \pm 0.2)$\,eV. See text for more details.}\label{tab:DB}
%~\cite{Doke:1988dp} 
\end{table*}

In summary, the empirical formula of Eq.\,\ref{eq:R-definition} is able to successfully model both S1 quenching and S2 yield for ER events between 50 and 500\,keV at $50 \leq \edrift \leq 1000$\,V/cm.
%demonstrating a non-negligible contribution from escape electrons for $\edrift < 200$\,V/cm. 
In addition, the combined scintillation and ionisation analysis validates the generally-accepted assumptions on $\alpha$ and $W_{ph}(\textrm{max})$ for \LAr\  in the energy range typically relevant for dark matter physics experiments.


\section{Conclusions} \label{sec:conclusions}
The \ReD\ experiment aims to investigate the directional sensitivity of argon-based \TPC s to nuclear recoils in the energy range of interest for WIMP dark matter searches. A compact double-phase argon \TPC, featuring innovative readout by cryogenic SiPMs, was recently constructed for \ReD\ and fully characterised using $\gamma$-ray and neutron sources. Measurements of the single-photoelectron response, single-photon resolution, S1 scintillation light yield, and duplication factor due to crosstalk and afterpulsing were periodically performed over more than five months of continuous operation and found to be reproducible to within 1--2\%, demonstrating stable operation of the SiPMs at cryogenic temperature and stability of the optical properties of the \TPC\ and wavelength-shifting (TPB) coating. The purity of the \LAr, maintained by a recirculation loop, results in an electron lifetime above 2\,ms, significantly longer than the maximum electron drift time in the \TPC\ at the operational $\edrift$.

\TPC\ performance criteria have been defined and evaluated in the context of the scientific goals of \ReD. The scintillation and ionisation gains, $g_1$ and $g_2$, were derived using measurements of S1 and S2 taken in single- or double-phase mode at $0 \leq \edrift \leq 1000$\,V/cm. The measured value of $g1$ is $\sim$24\% higher than that measured in DarkSide-50 and almost a factor of 2 higher than in SCENE, an improvement driven mostly by better optical coverage of the \ReD\ \TPC\ and higher detection efficiency of the SiPMs with respect to photomultipliers. The ionisation amplification of the \ReD\ \TPC\ is comparable to that of DarkSide-50 and more than a factor of 6 higher than that of SCENE. The dispersion in the ratio of the ionisation to scintillation signals (S2/S1) is found to be 12-13\% for nuclear recoils in the energy range 20--80\,keV$_{nr}$, improving on previous results obtained by SCENE. Based on the measured values of $g_1$, $g_2$, and the S2/S1 dispersion, the performance of the \ReD\ \TPC\ satisfies the requirements needed to achieve the scientific aim of the \ReD\ experiment. Finally, a phenomenological parameterisation of the recombination probability in \LAr\  has been applied in order to describe the scintillation and ionisation response of \ReD\ in a consistent framework. The parametrisation provides a good description of the dependence of scintillation quenching and charge yield on the drift field for energies between 50--500\,keV and fields up to 1000\,V/cm. 

%Data taken with \ReD\ in single- or double-phase modes corroborate the phenomenological parameterisation of the recombination probability in \LAr\ from ARIS~\cite{Agnes:2018mvl} and contribute by extending its range of validity to drift fields up to 1000\,V/cm. The parametrisation provides a good description of the dependence of scintillation quenching and charge yield on the drift field measured in this work, for energies between 50--500\,keV and for fields up to 1000\,V/cm. 

%To mention:
%\begin{itemize}
%\item long-term stable operation of SiPM at cryogenic temperature
%\item purity of LAr
%\item TPC performance in S1 (10 PE/keV) and stability
%\item S2/S1 for ER and NR
%\item g1 and g2
%\item recombination parametrization, to describe scintillation quenching and 
%\Qy\ for energy 50-500~keV and fields up to 1000 V/cm.
%\end{itemize}

\appendix

\section{S3 ``echo'' events} \label{sec:echoes}
\label{sec:S3appendix}
Similar to DarkSide-50~\cite{Agnes:2015ftt}, S3 ``echo'' events are observed in \ReD, produced when photons from an S2 signal hit the cathode and extract one or more additional electrons. These electrons are then transported by the drift field, extracted and eventually accelerated, producing a delayed electroluminescence signal (S3). Since they travel from the cathode, the delay with respect to S2 corresponds to $T_{max}$, defined in Sec.\,\ref{sec:drifttime}. S3 signals allow an independent measurement of the S2 gain ($g_2$), but since they originate from one or a few electrons, their amplitude is generally very small ($< 50$\,PE). Due to sub-optimal efficiency of the standard reconstruction algorithm in this regime, these data are reconstructed with a relaxed threshold. 

\begin{figure*}
\centering
\includegraphics[trim=0 0 1.4cm 0.5cm,clip,width=0.95\columnwidth]{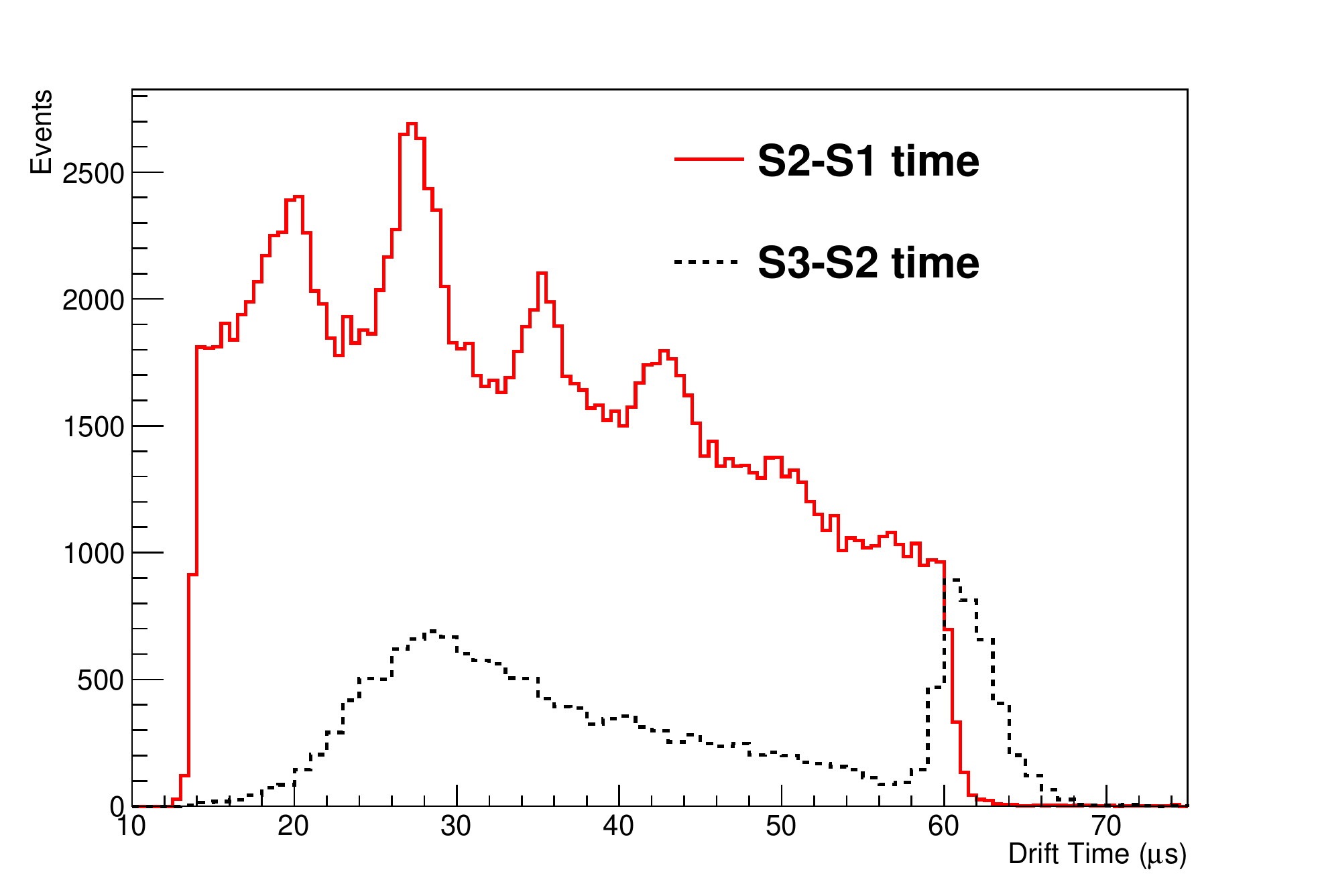}
\hspace{0.3cm}
\includegraphics[trim=0 0 1.4cm 0.5cm,clip,width=0.95\columnwidth]{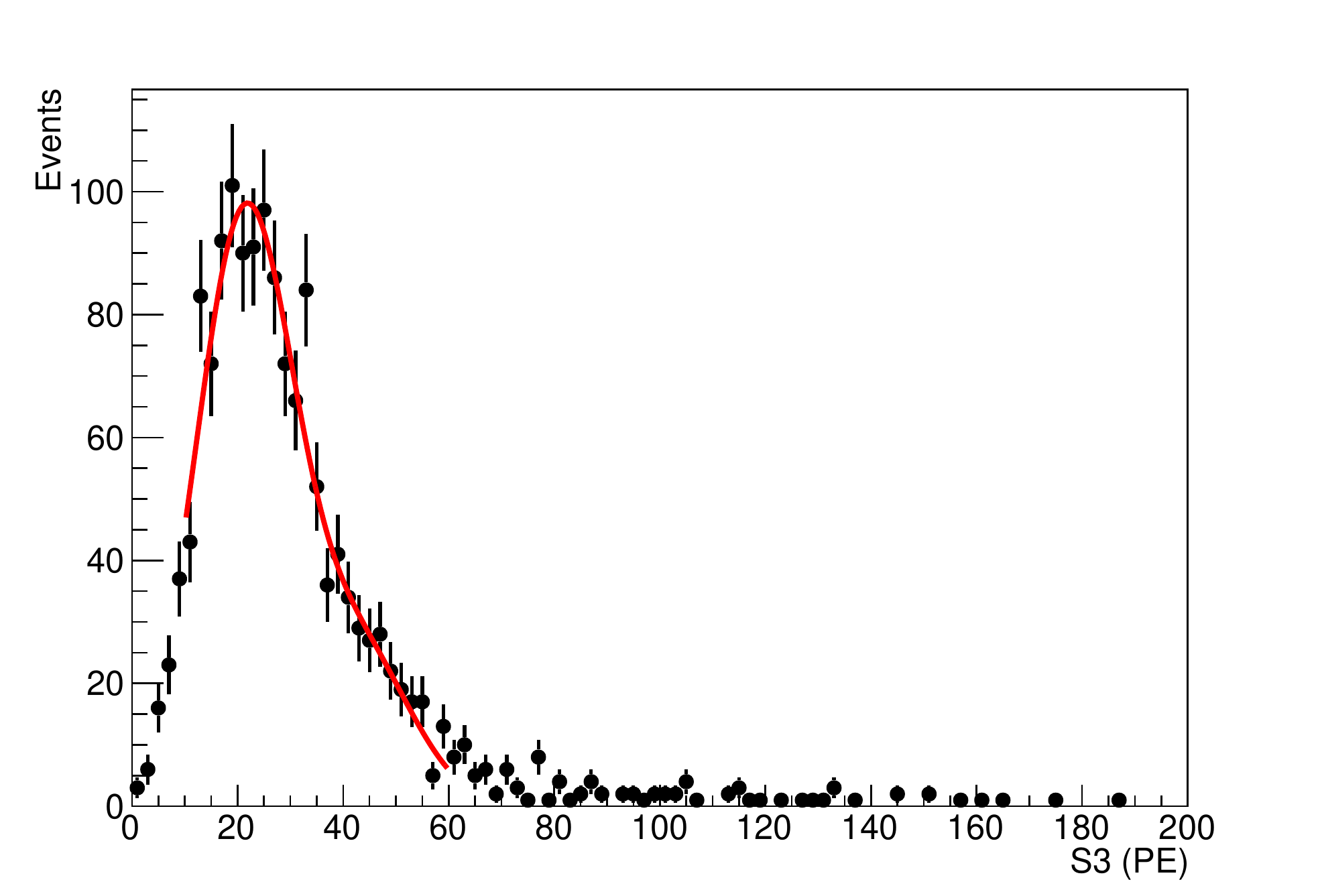}
\caption[Drift time distributions of S3, from several \Am\ calibrations]{\textit{Left:} Distribution of 
the drift time between S1 and S2 signals (red curve) and between S2 and S3 signals (black curve) for a set of \Am\ measurements
taken in double-phase mode at $\edrift = 183$\,V/cm. \textit{Right:} S3 charge distribution for events in which the delay 
between S2 and S3 signals is within $(T_{max} \pm 4)$\,\textmu s, superimposed with the fit described in the text.}
\label{fig:echoes}
\end{figure*}

Fig.\,\ref{fig:echoes} shows the drift time distribution between S1 and S2 events (red curve) and between S2 and S3 events (black curve), obtained from a set of \Am\ measurements at $\edrift = 183$\,V/cm. While the S2\,$-$\,S1 drift time distribution exhibits peaks and valleys caused by the presence of the copper field-shaping rings (see Sec.\,\ref{sec:drifttime}), the S3\,$-$\,S2 drift time distribution has a peak at approximately $T_{max}=62$\,\textmu s, thus providing evidence that these events mostly consist of ``echoes'' due to secondary photoionisation from the cathode. 
%The remainder of the distribution is likely due to late photoelectrons from S2 afterpulsing, combined with single electrons from background activity or impurities. 
The charge spectrum of the S3 events in the time window between 58 and 66\,\textmu s following the preceding S2 pulse and detected in the inner eight SiPMs of the top tile is shown in the right panel of Fig.\,\ref{fig:echoes}. 
Under the assumption that the S3 signals originate from one or two electrons, each contributing a signal of $g_2$ photoelectrons, the spectrum can be fitted with a sum of two Gaussian functions, with central values $g_2$ and $2g_2 $ and standard deviations $\sigma$ and $\sqrt{2}\sigma$. The best-fit curve is shown in Fig.\,\ref{fig:echoes} in red and corresponds to 
$g_2 = (21.0 \pm 0.8_{\rm \,stat}$)\,PE/e$^-$. The uncertainty due to the efficiency profile of these low-energy signals is evaluated by varying the lower bound of the fit range from 10 to 15\,PE, ultimately leading to a value of $g_2 = (21.0 \pm 1.3_{\rm \,stat+syst}$)\,PE/e$^-$, consistent with the value derived using the S1-S2 correlation discussed in Sec.\,\ref{sec:doke}.

\begin{acknowledgements}
M.~Leyton is supported by funding from the European Union's Horizon 2020 research and innovation programme under the Marie Sk\l odowska-Curie grant agreement No.\ 754496. The activity of V.~Oleynikov within this project has been supported by the SSORS funds of INFN. M.~Wada is supported by IRAP AstroCeNT funded by FNP from ERDF. 
The authors thank Fabrice Retiere (TRIUMF) and Roberto Santorelli (CIEMAT) for very useful comments on the manuscript.

\end{acknowledgements}

\bibliographystyle{spphys}       % APS-like style for physics
\bibliography{ReD_paper}   % name your BibTeX data base

%\begin{thebibliography}{}
% etc
%\end{thebibliography}

\end{document}